\documentclass{natureprintstyle}
\usepackage{epsfig,caption}
\usepackage{color}
\usepackage{bm}
\usepackage{graphicx}
\usepackage{longtable}
\usepackage{amssymb}
\usepackage{rotating,xcolor}
\usepackage{booktabs}

\usepackage{array}
\usepackage{float}
\usepackage{color}
\usepackage{amsmath,amssymb}
\usepackage{graphicx}
\usepackage[hidelinks]{hyperref}
\providecommand{\araa}{ARA\&A}%
\providecommand{\apj}{ApJ}%
\providecommand{\apjl}{ApJ}%
\providecommand{\apjs}{ApJS}%
\providecommand{\aap}{A\&A}%
\providecommand{\mnras}{MNRAS}%
\providecommand{\pasp}{PASP}%
\providecommand{\rmxaa}{Rev. Mexicana Astron. Astrofis.}%
\providecommand{\ssr}{Space~Sci.~Rev.}%
\usepackage[normalem]{ulem}

\def\afe{{\rm [}\alpha{\rm /Fe]}}
\usepackage{caption}
\DeclareCaptionLabelFormat{bold}{\textbf{#1~#2}.}

\newcommand{\useMainTextCaptions}{%
  \captionsetup[figure]{name=Figure, labelformat=bold, labelsep=space}%
  \captionsetup[table]{name=Table,  labelformat=bold, labelsep=space}%
  \providecommand{\figureautorefname}{Figure}%
  \providecommand{\tableautorefname}{Table}%
  \renewcommand{\figureautorefname}{Figure}%
  \renewcommand{\tableautorefname}{Table}%
}

\newcommand{\useExtendedDataCaptions}{%
  \setcounter{figure}{0}%
  \setcounter{table}{0}%

  \captionsetup[figure]{name={Extended Data Figure}, labelformat=bold, labelsep=space}%
  \captionsetup[table]{name={Extended Data Table},  labelformat=bold, labelsep=space}%

  \captionsetup[figure*]{name={Extended Data Figure}, labelformat=bold, labelsep=space}%
  \captionsetup[table*]{name={Extended Data Table},  labelformat=bold, labelsep=space}%

  \renewcommand{\figureautorefname}{Extended Data Figure}%
  \renewcommand{\tableautorefname}{Extended Data Table}%
}

\newcommand{\figref}[1]{%
  \hyperref[#1]{Figure~\ref*{#1}}%
}

\newcommand{\extfigref}[1]{%
  \hyperref[#1]{Extended Data Figure~\ref*{#1}}%
}

\newcommand{\tabref}[1]{%
  \hyperref[#1]{Table~\ref*{#1}}%
}

\newcommand{\exttabref}[1]{%
  \hyperref[#1]{Extended Data Table~\ref*{#1}}%
}

\newcommand\arcsec{\mbox{$^{\prime\prime}$}}%
\def\farcs{%
 \mbox{%
  \kern  0.13ex.%
  \kern -0.95ex\arcsec%
  \kern -0.1ex%
 } 
}

\usepackage{multirow}

\title{The Ashes of Supermassive Stars: \\
Globular Cluster-like Aluminum Enhancement in Little Red Dots}

\author{V. ~Kokorev$^{1,2,\dagger}$,
J. ~Chisholm$^{1,2}$, 
R.P. ~Naidu$^{3}$,
M. ~Gieles$^{4,5,6}$,
S. ~Finkelstein$^{1,2}$,
D. ~Berg$^{1,2}$,
H. ~Akins$^{7}$,
A. ~Taylor$^{1,2}$,
S. ~Fujimoto$^{8}$,
L.J. ~Furtak$^{1,2}$,
J. ~Greene$^{7}$,
A. ~de Graaff$^{9}$,
K. ~Hawkins$^{1,2}$,
T. ~Hsiao$^{1,2}$,
D. ~Nandal$^{10}$,
J. ~Matthee$^{11}$,
S. ~Monty $^{12,13}$,
P. ~Rinaldi $^{1,2}$,
M. ~Boylan-Kolchin $^{1,2}$
}

\begin{document}
\maketitle

\let\thefootnote\relax\footnote{

\begin{affiliations}
$\dagger$ NASA Hubble Fellow
\end{affiliations}
}

\setcounter{figure}{0}
\setcounter{table}{0}  
\useMainTextCaptions

\vspace{0.cm}
\begin{abstract}
    {\boldmath
The relative abundances of elements in galaxies serve as fossil records of the physical conditions and processes by which they were forged. While the Big Bang produced only the lightest elements, subsequent stellar nucleosynthesis imprinted characteristic abundance patterns onto the surrounding gas, set initially by the temperatures reached inside stars and subsequently shaped by how the processed material was mixed and released. Globular clusters -- dense, ancient groups of stars -- provide a striking unique example. Some contain stars depleted in magnesium and enriched in aluminum, showing that they formed from gas exposed to exceptionally hot hydrogen burning. The stars responsible remain unknown. Little Red Dots \cite{matthee23} may provide this missing engine. These compact, luminous objects formed at cosmic epochs similar to those associated with globular-cluster formation \cite{chisholm26,mbk18} and are enshrouded by dense gas \cite{inayoshi25,naidu25_bh*,degraaff25_cliff,rusakov25} whose chemical composition can be measured with the James Webb Space Telescope. Here, using deep spectroscopy from the SPURS program, we show that this abundance pattern characterizes the LRD central engine: magnesium-depleted and aluminum-enhanced gas with a metallicity only 1\% that of the Sun. This pattern is not produced by ordinary massive stars at these redshifts and cannot be mimicked by ionization, gas geometry or dust. Instead, it is reproduced by hot hydrogen burning in fully convective supermassive stars, with the measured abundances implying masses of at least 10,000 solar masses -- approximately 100 times larger than any star observed in the present-day Universe \cite{prantzos17,gieles18,gieles25}. Little Red Dots may therefore reveal supermassive stars during their brief lives or in the immediate aftermath of their direct collapse, simultaneously identifying the long-sought source of globular cluster abundance anomalies and a formation pathway for massive black hole seeds.
}

\end{abstract}

Little red dots (LRDs) \cite{matthee23} have emerged as one of the most enigmatic populations discovered in the early Universe. Initially interpreted as extraordinarily compact galaxies ($\lesssim40$ pc; unresolved in the rest-optical even with JWST \cite{furtak24}) that had assembled implausibly large stellar masses within the first billion years after the Big Bang \cite{labbe22,boylan-kolchin23}, this picture was soon challenged by the first JWST spectra. These observations revealed that LRDs have nearly ubiquitous broad emission lines from rapidly moving gas, resembling those seen around actively growing supermassive black holes \cite{matthee23,kokorev23c,furtak24,greene24}. 

Despite initial progress, the physical nature of LRDs remains uncertain. Early interpretations favored heavily obscured active galactic nuclei (AGN) \cite{furtak24,kokorev23c,greene24}, although the implied black-hole-to-host mass ratios are often an order of magnitude larger than those observed in the local Universe \cite{greene05,maiolino23b,pacucci24}. More recently, a growing body of evidence has suggested that the broad emission lines arise within exceptionally dense, gas-rich environments, capable of producing the extreme Balmer breaks, strong [Fe\,{\sc ii}] emission, and narrow P-Cygni absorption features that characterize the population \cite{inayoshi25,naidu25_bh*,degraaff25_cliff,rusakov25,torralba25,kokorev26,matthee26}. These models have begun to converge on the physical conditions surrounding the central source, however the nature of the central engine itself remains unresolved. Proposed explanations include rapidly growing supermassive black holes \cite{naidu26_lrds,madau26,kokorev26,liu26}, quasi-stars \cite{begelman26}, and supermassive stars \cite{chisholm26,nandal26,martins26}, illustrating that the observed continuum and emission-line properties do not yet uniquely identify the nature of the central engine.

\begin{figure*}[t!]
  \centering
  \includegraphics[width=\textwidth]{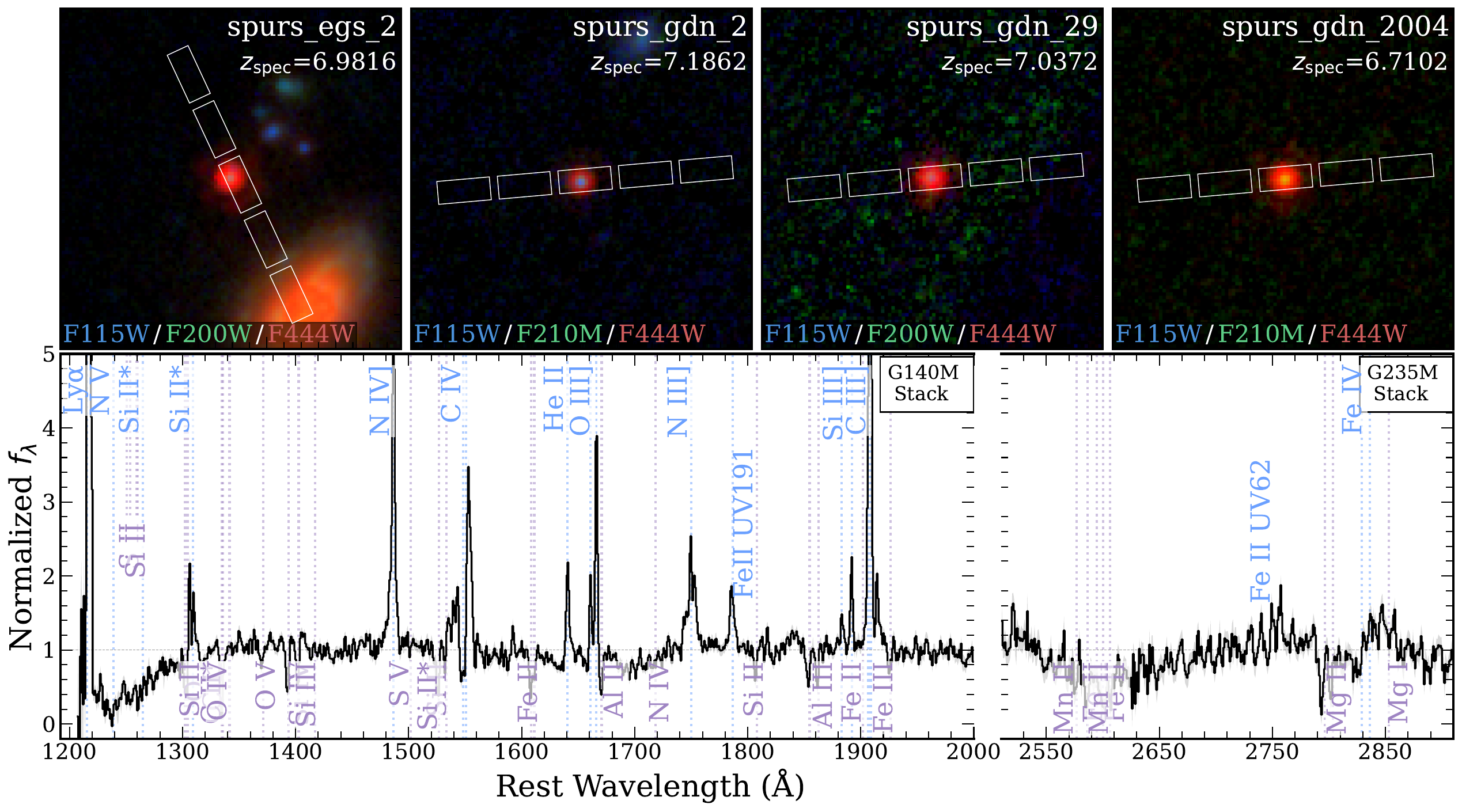}
  \includegraphics[width=\textwidth]{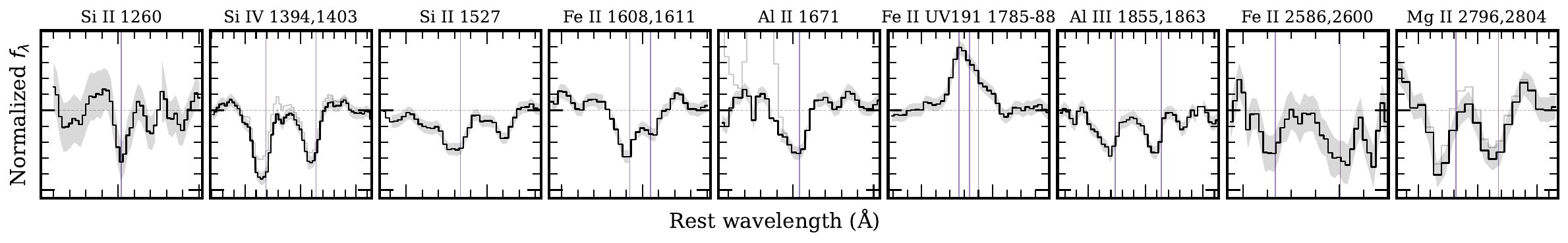}
  \caption{
\textbf{JWST imaging and rest-frame UV spectrum stack of the LRD sample.}
\textbf{Top:} JWST/NIRCam three-color composites (F115W, F200W/F210M and F444W) of the four little red dots (LRDs) analysed in this work. White outlines indicate the SPURS NIRSpec microshutter configuration. All sources are compact and exhibit the characteristic rest-optical dominance of the LRD population.
\textbf{Middle:} Inverse-variance-weighted stack of spectra of the four LRDs. The principal Mg, Al, Si and Fe absorption features used throughout this work are indicated. The depth of the stacked spectrum enables robust measurements of weak ultraviolet absorption lines from which the elemental abundances are derived.
\textbf{Bottom:} Zoom-in panels of absorption lines used to determine elemental abundances. Local continua have been modeled and subtracted to allow for robust line modeling. Where applicable the emission part of the P-Cygni has been subtracted. We also highlight some fluorescent Fe\,{\sc ii}
lines, now routinely identified in LRDs \cite{kokorev26}.}
\label{fig:fig1}
  \vspace{-0mm}
\end{figure*}

While the broad emission lines, breaks and continua might help in constraining the geometry and physical state of the gas, the elemental abundances of LRDs provide a fundamentally different probe. Their dense environments retain recently ejected material from the central source, allowing the surrounding gas to preserve a relatively direct record of the nucleosynthetic processes that enriched it for at most a few million years, before subsequent mixing and core-collapse supernova enrichment erase the signature. The metal absorption is kinematically coincident with the blueshifted P-Cygni absorption in broad H$\beta$, directly linking the absorbing gas to the dense material surrounding the central engine (\extfigref{ext_fig:fig3}).
The rest-frame ultraviolet is uniquely suited to this task. Strong Al, Mg and Si transitions are largely absent from the rest-frame optical, whereas the ultraviolet hosts numerous resonance lines of these elements together with Fe, enabling direct measurements of the chemical composition of the dense gas in LRDs. Here, using the unprecedented UV depth of the SPURS JWST/NIRSpec medium-grating spectroscopy \cite{chen26}, we measure ionic column densities from Mg\,{\sc ii}, Al\,{\sc iii}, Fe\,{\sc ii}, Si\,{\sc ii} and Si\,{\sc iv} absorption and infer the abundances of Mg, Al, Fe and Si for the first time in a sample of four $z\sim7$ LRDs (\figref{fig:fig1}). The robustness of these column-density measurements, including detailed tests for line saturation, is discussed in the Methods. These objects are among the brighter spectroscopically confirmed LRDs known at this epoch, enabling the detection of otherwise weak ultraviolet (UV) absorption features.

\begin{figure*}[t!]
  \centering
  \includegraphics[width=\textwidth]{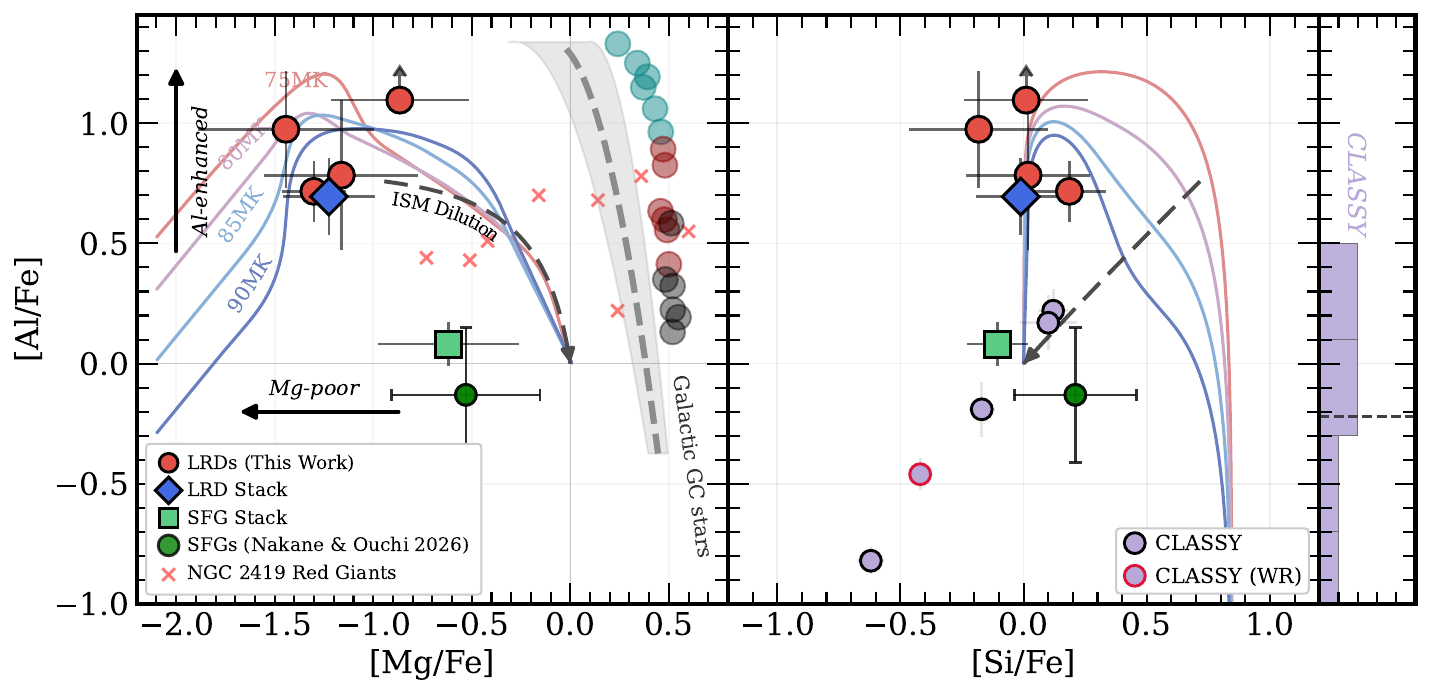}
 \caption{
\textbf{Al enhancement and Mg depletion in LRDs.}
\textbf{Left:} Relative abundances of Mg and Al with respect to Fe for the individual LRDs (red circles) and their inverse-variance-weighted stack (blue diamond). For comparison with globular clusters, colored circles show NGC 6752 stars grouped by light-element enrichment: unenriched (black), moderate (dark red) and high (teal) \cite{yong05,bastian18}; the gray shaded region shows Galactic globular-cluster stars \cite{milone22}, and red crosses show NGC 2419 red giants exhibiting some of the most extreme Mg depletion and Al enhancement known \cite{cohen_kirby12}. The LRDs occupy the extreme Al-rich, Mg-poor region of this globular-cluster abundance sequence. The contrast with star-forming galaxies is dramatic. The matched SPURS star-forming-galaxy stack (green square) has both substantially lower [Al/Fe] and higher [Mg/Fe] than the LRDs; the independent Al abundance measured in a single SPURS galaxy is also shown \cite{nakane26}. The LRDs are therefore strongly offset from star-forming galaxies at similar redshift and UV luminosity in both abundance ratios.Colored curves show hydrogen-burning calculations at different temperatures. The LRD abundances lie close to the most strongly processed material produced by high-temperature proton-capture burning; dilution of this material with ambient gas moves the abundances along the gray dashed curve toward those observed in globular-cluster stars.
\textbf{Right:} Relative abundances of Si and Al with respect to Fe. Purple circles and the histogram show measurements from the local CLASSY galaxy sample \cite{classy_main} for comparison. The LRDs are similarly strongly enhanced in [Al/Fe] relative to local star-forming galaxies, while retaining approximately Solar [Si/Fe].
}
\vspace{-6mm}
\label{fig:fig2}
\end{figure*}

\begin{figure*}
  \centering
\includegraphics[width=1.\linewidth]{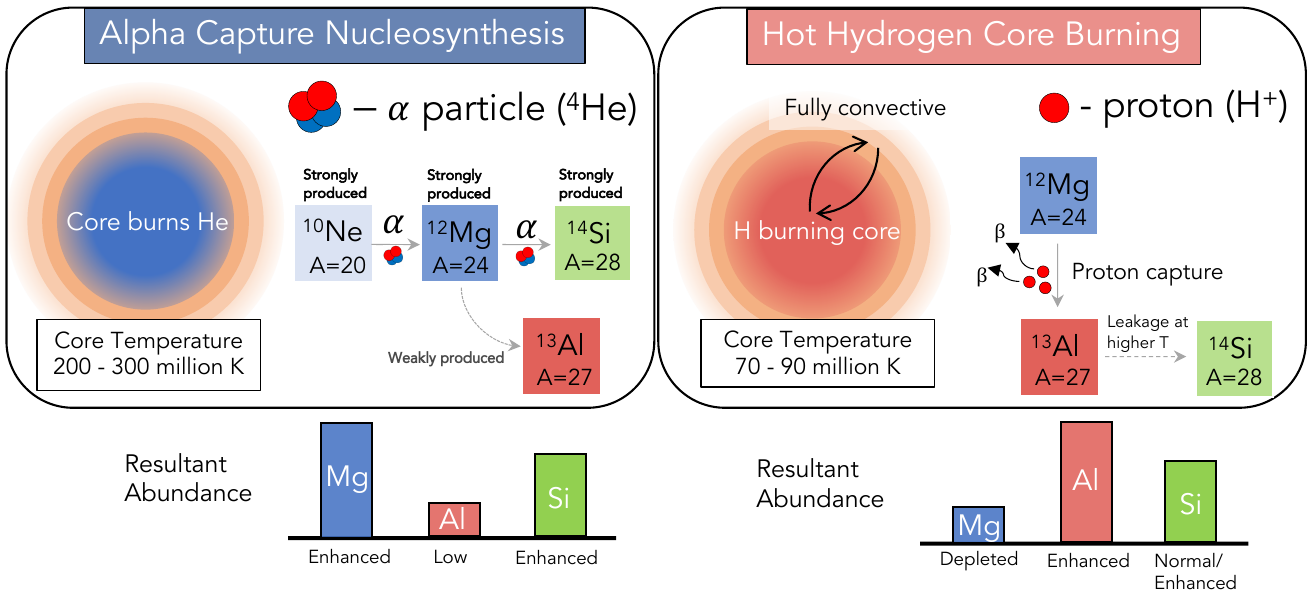}
\caption{\textbf{Abundance patterns retain the signatures of nucleosynthesis.}
\textbf{Left:} Successive captures of alpha particles ($^4$He) preferentially synthesize the even-atomic-number elements Mg and Si, 
while the production of odd Al is low. \textbf{Right:} 
In fully convective stars undergoing hot hydrogen burning, core temperatures can reach 70 -- 90 million K, enabling proton capture and conversion of Mg directly into Al. This produces the contrasting signature of Mg depletion, strong Al enhancement and normal (solar) or moderately enhanced Si.}
\label{fig:fig3}
\vspace{-6mm}
\end{figure*}

The LRDs we report here exhibit a highly unusual abundance pattern: strong magnesium depletion and aluminum enhancement, while silicon remains close to the Solar abundance relative to iron (\figref{fig:fig2}). Magnesium is depleted by approximately an order of magnitude relative to the Solar [Mg/Fe] ratio, whereas aluminum is enhanced by $\gtrsim0.5$ dex. From the damped Ly$\alpha$ absorption, we also measure a low metallicity of [Fe/H]$\sim-2.5$. This combination is difficult to reconcile with conventional galactic chemical enrichment. Core-collapse and pair-instability supernovae broadly synthesize Mg, Si and Fe together \cite{kobayashi20}, while alpha-capture nucleosynthesis proceeds through successive captures of $^4$He nuclei and therefore preferentially produces the even-atomic-number elements Mg and Si rather than the odd-atomic-number element Al. The observed pattern instead points to proton-capture nucleosynthesis during hydrogen burning at high temperatures, in which magnesium is converted directly into aluminum during hot hydrogen burning. Among the light-element abundance anomalies observed in globular clusters (GCs) -- dense, ancient stellar systems thought to have formed rapidly in the early Universe \cite{bastian18,milone22} -- the Mg--Al anticorrelation is one of the most temperature-demanding ($T_{\rm core}\sim{\gtrsim70}$ MK \cite{prantzos17}) because it requires proton captures to reach heavier nuclei than those involved in the carbon--nitrogen--oxygen (CNO) cycle ($T_{\rm core}\sim {\gtrsim20}$ MK). At still higher temperatures ($\gtrsim80$ MK), leakage from the Mg--Al cycle enhances silicon, while the rare Mg--K anticorrelation, seen only in the most massive GCs (e.g. NGC 2419 red giants shown in \figref{fig:fig2} \cite{cohen_kirby12,mucciarelli15}), requires even more extreme burning \cite{iliadis16,prantzos17}. The observed combination of strong magnesium depletion, aluminum enhancement and little silicon enrichment therefore confines the burning temperature to approximately 73--81~MK. We illustrate the contrasting abundance patterns produced by alpha capture and hot hydrogen burning nucleosynthetic pathways in \figref{fig:fig3}.

Such an Al-rich, Mg-poor signature is highly unusual outside GCs. Although anomalous light-element abundances -- most commonly nitrogen enrichment -- are widespread among GC stars and are also found in field populations thought to originate in now-disrupted clusters \cite{schiavon17,belokurov24}, the extreme combination of magnesium depletion and aluminum enhancement observed here is much rarer. This pattern has so far been observed only in the oldest, most massive and most metal-poor GCs \cite{fernandez-trincado17,naidu22_gc}, where it requires proton-capture burning at $T_{\rm core}\gtrsim70$ MK. No secure non-cluster analog of this extreme abundance pattern is known. We therefore searched for comparable aluminum enhancement in galaxies.

We do not find a comparable abundance pattern in either extreme local star-forming galaxies or in ordinary galaxies matched to the LRDs in redshift and UV luminosity. At low redshift ($z<0.2$), we compare to the COS Legacy Archive Spectroscopic SurveY \cite{classy_main}, a deep ultraviolet spectroscopic atlas of 45 galaxies spanning $6.2 < \log M_\star (M_\odot) < 10.1$, with star-formation rates enhanced by $\sim2$ dex over the $z\sim0$ star-forming main sequence and therefore more representative of the intense star formation seen at $z>2$. No CLASSY galaxy with secure aluminum abundance measurement (SNR$>3$), shows a similar aluminum enhancement.

Likewise, an identically processed SPURS stack of star-forming galaxies matched to the LRDs in redshift and UV luminosity shows a strikingly different absorption pattern (\extfigref{ext_fig:fig2}). Al\,{\sc iii} is substantially stronger in the LRDs, while Mg\,{\sc ii} is comparable to or weaker than in the matched galaxies and Fe\,{\sc ii} in LRDs is enhanced only moderately. At the same time the [Si/Fe] remain broadly consistent between the two populations. The aluminum enhancement is therefore visible directly in the spectra relative to both Mg, Fe and Si, independent of the detailed conversion from absorption strength to column density. 

The close resemblance to the abundance anomalies of globular clusters suggests that the gas in both environments was processed under similarly extreme hydrogen-burning conditions. Notably, our sample comprises some of the brightest LRDs known at $z\sim7$, while similarly extreme Mg-poor, Al-rich abundance patterns in the local Universe are observed only in the oldest, most massive and most metal-poor GCs. Producing and releasing such material is, however, difficult within ordinary stellar evolution. To preserve the Mg-poor, Al-rich signature, material processed in the hot hydrogen-burning core must be continuously transported outward and replaced by fresh hydrogen before later burning stages alter its composition \cite{prantzos17,gieles18}. Massive stars, tens to a hundred times the mass of the Sun, can attain the required temperatures in the late stages of core hydrogen burning when the helium abundance is already too high to explain globular cluster abundances \cite{bastian18}, and the products remain confined to their convective cores and buried beneath largely unprocessed radiative envelopes. By the time stellar winds expose these inner layers during the Wolf--Rayet (WR) phase, the alpha-process has already modified the hot hydrogen-burning abundance pattern.

Asymptotic giant branch (AGB) stars provide another possible route through hot-bottom burning, in which the base of the convective envelope reaches the hydrogen-burning region \cite{karakas14}. At $z\sim7$, when the Universe is only 800 million years old, only the most massive AGB and super-AGB stars evolve rapidly enough to be relevant, requiring a pre-existing stellar population tens to hundreds of millions of years old. Such a scenario would therefore require a substantial earlier generation of star formation in addition to the compact, luminous source observed in the LRDs. Moreover, low-metallicity models of precisely these stars still fail to reproduce the extreme magnesium depletion accompanied by strong aluminum enhancement observed here \cite{karakas10,doherty14}. The source must therefore combine exceptionally hot hydrogen burning with efficient mixing throughout a hydrogen-rich stellar core, releasing the processed material before later burning stages erase its distinctive chemical signature.

\begin{figure*}
  \centering
\includegraphics[width=1.\linewidth]{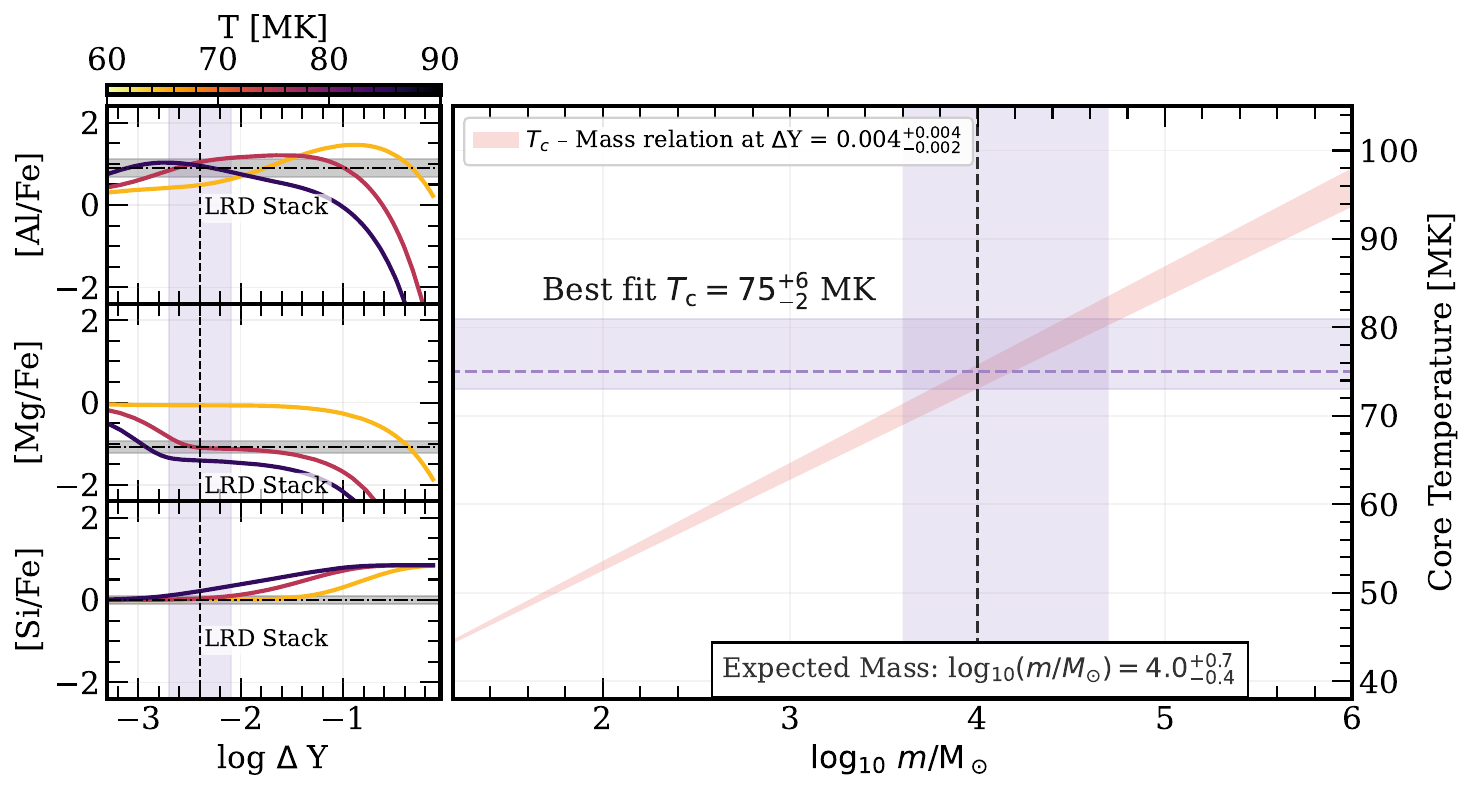}
\caption{\textbf{Hydrogen-burning abundance patterns and the implied mass of the polluting star.}
\textbf{Left:} Predicted evolution of the relative [Al/Fe], [Mg/Fe] and [Si/Fe] abundance patterns as a function of helium enrichment ($\Delta$Y). Curves represent solar-like $\alpha$ enhancement at [Fe/H] $= -2.5$ from nucleosynthesis calculations, colored by the temperature of the core ($T_{\rm c}$). Dashed horizontal bands indicate the abundances measured from LRD spectra, while the vertical line marks the degree of hydrogen processing that simultaneously reproduces the observed Mg depletion and Al enhancement without substantial silicon production.  \textbf{Right:} Central temperature as a function of stellar mass for a fully convective star at ($\Delta {\rm Y}=0.004^{+0.004}_{-0.002}$), following the fit from \cite{gieles25} to the stellar models of \cite{prantzos17}. The horizontal shaded region shows the temperature range inferred from the LRD abundances, and the vertical shaded region marks the corresponding stellar-mass range.}
\label{fig:fig4}
\vspace{-6mm}
\end{figure*}

Supermassive stars -- predicted by stellar-structure models but never directly observed, and previously invoked primarily through their ability to explain the anomalous abundance patterns of globular clusters -- could naturally satisfy these conditions \cite{denissenkov14,gieles18,gieles25}. Their internal structure depends on mass, accretion history and evolutionary stage \cite{hosokawa12,nandal26}, but models with sufficiently efficient mixing can expose material processed by hot hydrogen burning before later nucleosynthesis erases the signature. Comparing our measurements with the fully convective supermassive-star models considered here, we find excellent agreement with hydrogen-burning calculations at core temperatures of $\sim75$ MK, metallicities of [Fe/H] $\sim-2.5$, and helium enrichment of $\Delta Y\simeq0.004$, as shown by the colored curves in \figref{fig:fig2}. For a fully convective stellar structure, these conditions correspond to characteristic masses of $\sim10^{3.6 - 4.7}\,M_{\odot}$ \cite{prantzos17}. We show how the elemental abundances help us constrain the temperature and the mass of the central LRD engine in \figref{fig:fig4}. Independent support for this mass scale comes from recent work showing that the luminosities, pseudo-photospheres and P-Cygni winds of LRDs can be reproduced by interaction-powered eruptions from massive stellar engines, with inferred central masses of a similar order \cite{liu26,naidu26_lrds}. The abundance pattern reported here provides independent empirical evidence for the supermassive-star interpretation, requiring such a star to have been present within the past $\lesssim3$\,Myr. Stars of this mass are not expected to follow ordinary late-stage stellar evolution. Once they exhaust their hydrogen fuel and become unstable, they are instead predicted to encounter the general-relativistic instability and collapse directly into black holes \cite{nandal26_puls}, leaving massive remnants whose final masses depend on the amount of material lost during the preceding evolution \cite{nandal26_puls,hosokawa12}. Secure intermediate-mass black holes remain rare in present-day globular clusters, although one of the strongest cases is found in $\omega$ Centauri, where stellar kinematics imply a central black hole of at least $\sim8,000\,M_\odot$ \cite{haberle24}. Intriguingly, this mass scale is comparable to the characteristic masses inferred here for the polluting stars from the LRD abundance patterns. Supermassive-star formation in these systems may therefore provide a natural link between globular-cluster enrichment and the formation of massive black-hole seeds, although the final remnant mass depends on the subsequent growth, mass loss and evolution of the polluting star.

The kinematics provide an additional clue that the chemically processed material is being released from the LRD central source. The Al\,{\sc iii}, Mg\,{\sc ii} and other metal absorption lines are kinematically coincident with the absorption component of the broad H$\beta$ profile (\extfigref{ext_fig:fig3}), which traces the dense gas surrounding the LRD central source \cite{matthee26}. The same narrow absorption component is extremely metal poor, with ${\rm [Fe/H]}\sim-2.5$ inferred from the Fe\,{\sc ii} and neutral-hydrogen columns. In the supermassive-star interpretation, this close correspondence suggests that we are observing chemically processed material in the dense wind of the polluting star itself. This is precisely the material that can subsequently mix with the ambient gas and imprint the Mg-poor, Al-rich abundance pattern onto a forming globular-cluster population.

Together, the chemical and kinematic similarities, combined with the expectation that globular clusters underwent their initial formation at these same early cosmic epochs \cite{mbk18,chisholm26}, raise the possibility that LRDs are witnessing the same physical process invoked to explain their abundance anomalies. Models developed independently for globular-cluster formation predict that rapid gas inflow and stellar collisions can produce fully convective supermassive stars with masses approaching $\sim10^{4}\,M_{\odot}$ within $\sim1$ Myr \cite{gieles18,gieles25}, remarkably similar to the mass scale inferred here from the abundances alone. The agreement extends to metallicity. The Mg--Al anticorrelation reaches its most extreme form in the most metal-poor globular clusters \cite{fernandez-trincado17,cohen_kirby12}, and the ${\rm [Fe/H]}=-2.56\pm0.30$ measured for the narrow absorbing gas places the LRDs in the same metallicity regime. The abundance pattern, the inferred mass of the polluting star and the metallicity of the gas therefore converge on the same physical picture. The extreme gas densities, low metallicity and rapid inflows inferred for LRDs may thus provide precisely the conditions in which a compact stellar system assembles while a central object, or objects, grow to supermassive scales through sustained accretion and stellar collisions \cite{gieles25,chisholm26}.

In this picture, during hydrogen burning, convection and other mixing processes transport Mg-poor, Al-rich material from the stellar interior toward the surface, where winds and pulsational mass loss inject it into the surrounding gas. Strikingly, the metal absorption observed in the LRDs is kinematically coincident with the blueshifted absorption in the broad H$\beta$ profile (\extfigref{ext_fig:fig3}), directly linking the anomalous abundances to the dense outflow surrounding the central source. The LRDs may therefore reveal both sides of the enrichment process: the production of the Mg-poor, Al-rich material through hot hydrogen burning and its subsequent release into the surrounding environment. Substantial mass loss is intrinsic to this process, such that the initial mass of the supermassive star need not map directly onto the mass of any surviving black-hole remnant.

Once expelled, mixing of this processed material with ambient gas would dilute the extreme LRD abundance pattern toward that observed in present-day globular-cluster stars, following the gray dashed track in \figref{fig:fig2}. Stars forming from this mixture would inherit the characteristic light-element anomalies. This enrichment must occur rapidly, however: within a few million years, core-collapse supernovae from ordinary massive stars begin injecting newly synthesized Mg, Si, Fe and other elements, progressively erasing the proton-capture signature.

The exceptionally Mg-poor and Al-rich gas observed in the central engines of this sample of LRDs therefore constrains not only the mass of the polluting source, but also the short interval over which its chemical imprint remains visible. We are likely observing these systems during or shortly after the enrichment phase, while supermassive-star material is being released into the surrounding gas and before core-collapse supernovae from other massive stars substantially dilute the signature. 
Indeed, the extreme abundances suggest relatively little dilution, approaching the composition of nearly pure supermassive-star ``ashes''. For this abundance pattern to become imprinted in a globular-cluster stellar population, a new generation of stars must therefore form from this enriched material before the onset of substantial supernova enrichment. The short timescale also makes it plausible that the polluting supermassive star is still present; alternatively, it may have already undergone direct gravitational collapse into a massive black-hole seed \cite{nandal26_puls,chandrasekhar64,nagele24}. In the latter case, powering the observed LRD luminosities with a remnant of order $\sim10^4\,M_{\odot}$ would require very high, potentially super-Eddington accretion rates. Thus, LRDs may capture a brief stage in which supermassive-star enrichment, globular-cluster assembly and the formation and early growth of a massive black-hole seed are closely linked.

\setcounter{figure}{0}
\setcounter{table}{0}  
\useExtendedDataCaptions

\bibliography{refs}
\clearpage

\begin{methods}
\vspace{0.4cm}

In this paper, uncertainties are quoted at the $1\sigma$ level or as 68\% confidence intervals. Upper limits are reported at the $1\sigma$ level unless otherwise stated. We assume a flat $\Lambda$CDM cosmology with $\Omega_{\mathrm{m},0}=0.3$, $\Omega_{\mathrm{\Lambda},0}=0.7$, and $H_0=70$ km s$^{-1}$ Mpc$^{-1}$, together with a Chabrier \cite{meth:chabrier} initial mass function spanning $0.1$--$100\,M_{\odot}$. All magnitudes are expressed in the AB system \cite{meth:oke74}.

\noindent
{\bf Observations and Data Reduction}
\\
\noindent The grating spectra analyzed in this work were obtained as part of the JWST Cycle~4 SPURS program (Tang et al. in prep. \cite{meth:chen26}; PIs: C. Mason and D. Stark). SPURS provides $30$ h of G140M spectroscopy per MSA mask, enabling the rest-frame ultraviolet spectra of little red dots (LRDs) to be studied at unprecedented depth and spectral resolution. These observations are complemented by integrations of $8$ h with G235M and $3$ h with G395M, which provide robust spectroscopic redshifts and confirm the LRD nature of the targets through their broad hydrogen emission lines. The available observations span the EGS, Abell 2744, GOODS-N and GOODS-S fields.

We reduce these data using a custom wrapper around the JWST Data Calibration Pipeline v1.20.2. Briefly, we execute the \texttt{detector1} stage of the pipeline using the default \texttt{1481.pmap} parameters. We then run a custom routine to further remove the effects of 1/f noise and the ``picture frame'' effect from individual exposures (see H. Akins in prep. for details). We then run the \texttt{spec2} and \texttt{spec3} pipeline modules with default parameters to produce 2D spectra. We next perform an optimal extraction\cite{meth:horne86} to produce the final 1D spectra. 

For the accompanying imaging analysis, we homogeneously process the corresponding JWST/NIRCam observations with the \texttt{grizli} pipeline \cite{meth:grizli}. The reduction procedure is described in detail by \cite{meth:valentino23,meth:kokorev24} and within the DAWN JWST Archive (DJA). Photometric catalogs are constructed following the procedure outlined in \cite{meth:kokorev24,meth:kokorev25a}.

\noindent
{\bf Spectroscopic Redshift}
\\
\noindent
We derive spectroscopic redshifts from the SPURS grating spectra using a modified version of \texttt{msaexp} \cite{meth:msaexp,meth:kokorev24b}. The spectra are modeled with a flexible continuum represented by a series of cubic splines, together with Gaussian profiles for the emission lines. The centroid, amplitude, and width of each line are allowed to vary freely. For H$\beta$ and H$\alpha$, we additionally allow a two-component decomposition, requiring the narrow component to have $\mathrm{FWHM}<800$\,km\,s$^{-1}$ and the broad component to have $\mathrm{FWHM}>800$\,km\,s$^{-1}$. The redshift solutions are anchored to the bright, narrow forbidden lines, primarily the [O\,{\sc iii}]\,$\lambda\lambda4959,5007$ doublet. In total, we fit 298 objects, of which 263 yield secure redshift solutions, defined by the detection of at least two emission lines at ${\rm S/N}>3$. Throughout this work, we adopt the centroid of the [O\,{\sc iii}] doublet as the systemic redshift against which the velocity offsets of all absorption features are measured.

\noindent
{\bf Sample}
\\
\noindent We begin with the spectroscopically confirmed SPURS LRD sample. The LRDs are identified by using the now-standard criteria of a red rest-optical continuum, a blue rest-ultraviolet slope, and a compact morphology in the long-wavelength NIRCam bands \cite{meth:matthee23,meth:labbe23,meth:kokorev24,meth:kocevski24,meth:akins24}. We additionally require a broad Balmer-line component with $\mathrm{FWHM}>800$,km,s$^{-1}$, a feature now observed routinely throughout the spectroscopically confirmed LRD population \cite{meth:kocevski23,meth:greene24,meth:kokorev23c,meth:labbe24,meth:taylor25,meth:naidu25_bh*,meth:hviding25,meth:degraaff25_cliff}. Because our abundance analysis requires simultaneous coverage of the principal Mg, Al, Si, and Fe absorption features, we further restrict the sample to $z>6$, where all of these transitions fall within the NIRSpec grating wavelength coverage. These criteria identify four LRDs. 

Three sources were previously reported in the literature. \texttt{spurs\_egs\_2} was first identified in the RUBIES JWST spectroscopy as RUBIES-55604 \cite{meth:wang24}, and also as one of the ``Universe breakers'' \cite{labbe22}. In addition, \texttt{spurs\_gdn\_2} and \texttt{spurs\_gdn\_29} were identified as ID2756 and ID9094 in the FRESCO \cite{meth:oesch23} grism data \cite{meth:xiao23,meth:xiao25}. We adopt the original discovery names for these objects throughout the paper.

Reliable measurements of weak rest-frame ultraviolet absorption features also require a significant continuum detection. We therefore require a binned continuum ($\times2$) ${\rm S/N}\geq5$ in line-free regions adjacent to the principal Mg, Al, Si, and Fe transitions. All four LRDs satisfy this requirement and constitute our final sample. The sources span $z=6.7$--$7.2$, with a mean continuum ${\rm S/N}\simeq5$ per pixel (approximately $7$ per resolution element) and ultraviolet magnitudes of $M_{\rm UV}\approx-20.0\pm0.5$. All four sources are unlensed. The target properties are listed in \autoref{tab:lrd_sample_tab}, and their location relative to previously identified photometric \cite{meth:kokorev24,meth:rinaldi26} and spectroscopic LRD samples \cite{meth:de_graaff25} is shown in \extfigref{ext_fig:fig1}.

\begin{table}
\centering
\small
\caption{LRD Sample}
\setlength{\tabcolsep}{1pt}
\label{tab:sources}
\begin{tabular}{lccccc}
\hline\hline
SPURS ID & Lit. Name  & R.A. (deg) & Dec. (deg) & $z_{\rm spec}$$^{1}$ & $M_{\rm UV}$ \\
\hline
egs\_2   & RUBIES-55604\cite{meth:wang24} &  214.9830 & 52.9560 & 6.9816 & -19.83 \\
gdn\_2 & FRESCO-2756\cite{meth:xiao25} & 189.0835 & 62.2026 & 7.1862 & -20.44 \\
gdn\_29  &  FRESCO-9094\cite{meth:xiao25} & 189.0192 & 62.2435 & 7.0372 & -20.23 \\
gdn\_2004  & -- &  189.0325 & 62.2164 & 6.7102 & -20.66 \\
\hline
\label{tab:lrd_sample_tab}
\end{tabular}
\smallskip
\par\footnotesize 1. Measured from the [O\,\textsc{iii}]$\lambda \lambda 4959,5007$.
\end{table}

\begin{figure}[t!]
  \centering
\includegraphics[width=.95\linewidth]{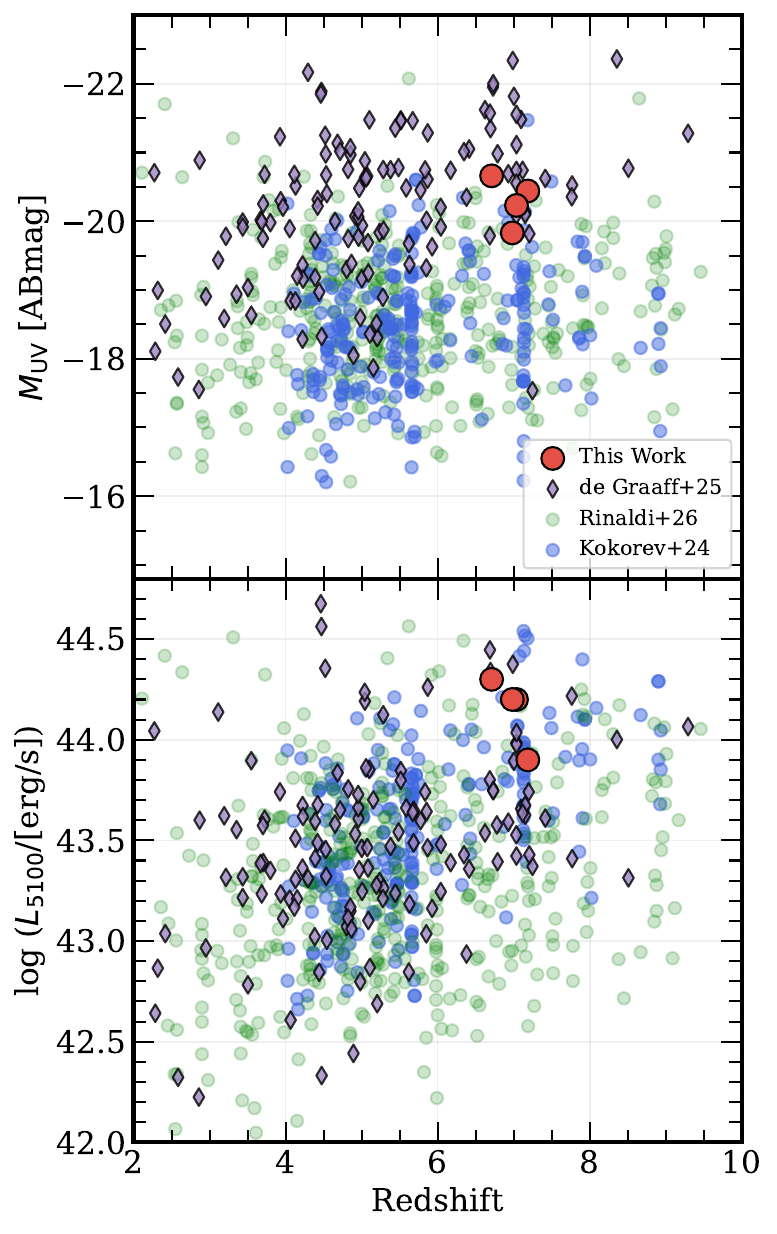}
\caption{\textbf{Rest-frame UV and optical properties of the LRD sample.}
UV \textbf{top}) and $\lambda_{\rm rest}=5100\,$ \AA\, \textbf{bottom}) luminosities against redshift of the LRDs analyzed in this work compared with previously reported photometric samples \protect\cite{meth:kokorev24,meth:rinaldi26} and spectroscopically confirmed LRDs \protect\cite{meth:de_graaff25}. The objects are some of the optically brightest LRDs at $z\sim7$, with  median $L_{5100}\sim 10^{44.2}$ erg/s.}
\label{ext_fig:fig1}
\end{figure}

\begin{figure*}
  \centering
\includegraphics[width=.7\linewidth]{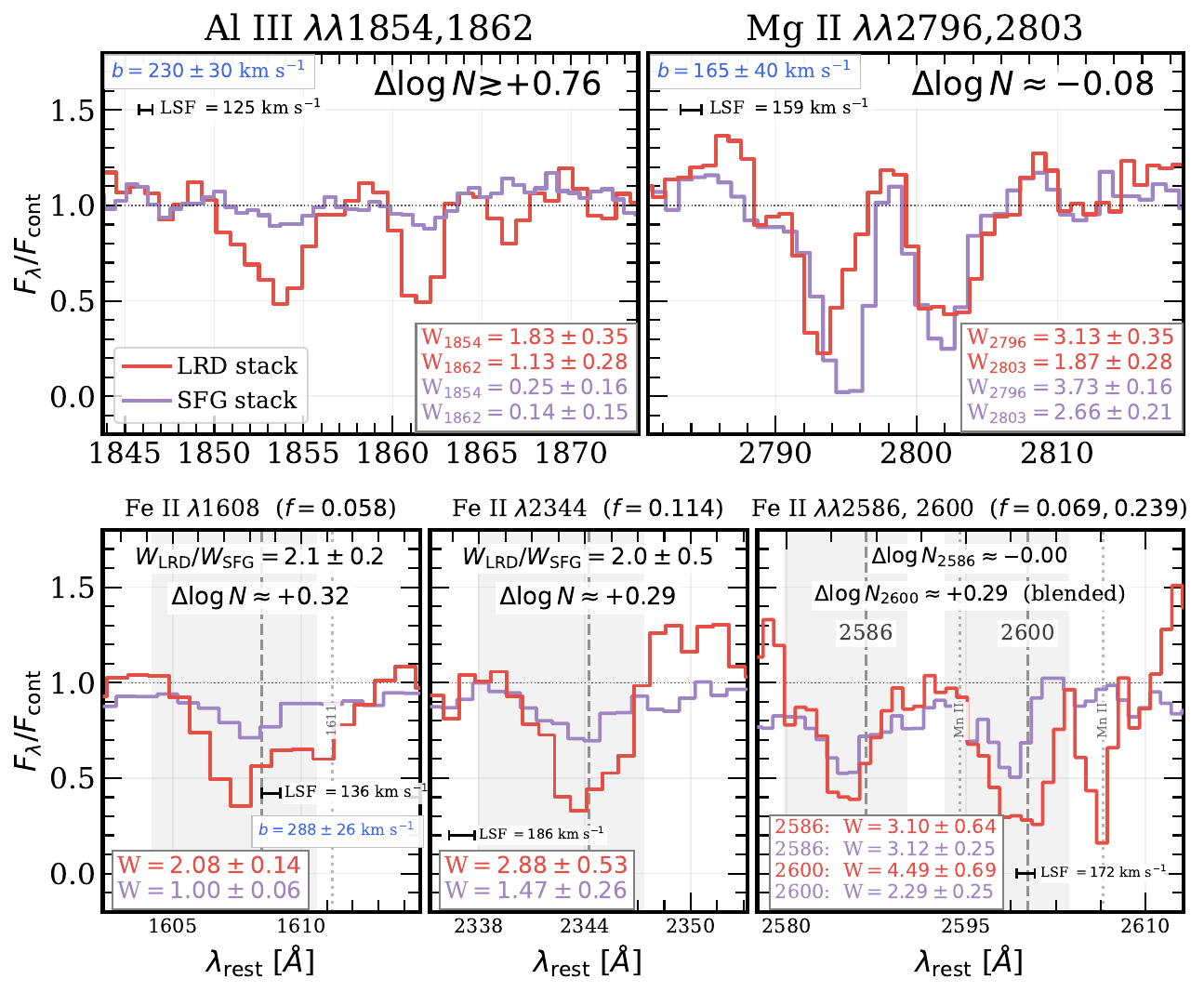}
\caption{\textbf{Direct comparison of Al, Mg, and Fe absorption in LRDs and SFGs.} Continuum-normalized absorption profiles of the inverse-variance-weighted LRD stack (red) and matched star-forming-galaxy stack (purple); oscillator strengths ($f$) of Fe\,{\sc ii} lines are listed in each panel title for comparison. \textbf{Top:} Al\,{\sc iii}\,$\lambda\lambda1854,1862$ is strongly detected in the LRDs but not significantly detected in the SFG stack (SNR$<3$), with at least an order-of-magnitude larger equivalent width. Mg\,{\sc ii}\,$\lambda\lambda2796,2803$ is comparable to or weaker than in the SFGs. The Al\,{\sc iii} doublet ratio remains inconsistent with strong saturation. \textbf{Bottom:} Fe\,{\sc ii}\,$\lambda1608$, $\lambda2344$, $\lambda2586$ and $\lambda2600$ differ much less between the two populations, demonstrating that the strong Al enhancement is present relative to both Fe and Mg. Intrinsic $b$ parameters and the instrumental LSF widths are also shown, highlighting that the absorption lines are resolved.}
\label{ext_fig:fig2}
\end{figure*}

\begin{figure*}
  \centering

\includegraphics[width=1.0\linewidth]{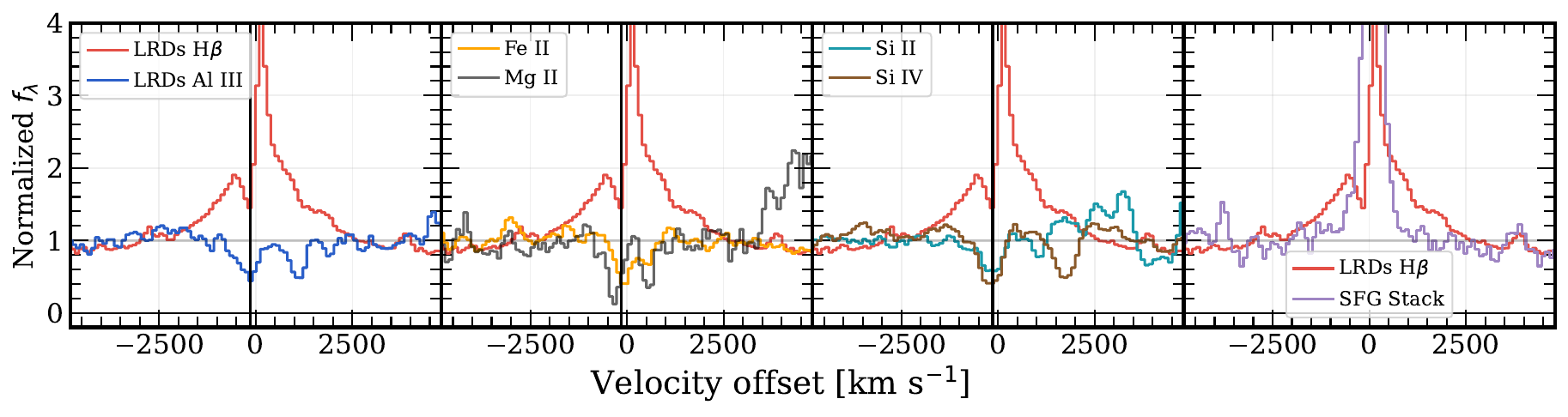}
\caption{\textbf{Broad-line emission and aluminum absorption in the LRD stack.}
Continuum-normalized line profiles from the stacked LRD spectrum, shown on a common velocity scale relative to the systemic redshift defined by [O\,{\sc iii}]. From left to right, the broad H$\beta$ profile in the LRD stack is compared with Al\,{\sc iii} $\lambda1855$, Fe\,{\sc ii} $\lambda1608$, Mg\,{\sc ii} $\lambda2796$, and Si\,{\sc ii} $\lambda1527$ and Si\,{\sc iv} $\lambda1394$. The blueshifted absorption in the H$\beta$ is kinematically coincident with all absorption features at $\sim-140\,\mathrm{km\,s^{-1}}$ away from the systemic redshift. (vertical line). This demonstrates that aluminum-rich and magnesium-poor gas lies around the central engine, rather than the LRD host galaxy. The rightmost panel compares H$\beta$ in the LRD and matched star-forming-galaxy stacks, the latter lacks both the broad wings and blueshifted absorption characteristic of the LRD spectrum. Profiles have been rescaled vertically for visual comparison, but no velocity shifts have been applied.}
\label{ext_fig:fig3}
\end{figure*}

\noindent
{\bf Spectral Stacking}
\\
\noindent
To increase the continuum ${\rm S/N}$ and characterize the average properties of the LRD population, we construct composite spectra from the four sources described above. Although stacking such a small sample can bias the composite toward individual objects, it substantially improves the continuum ${\rm S/N}$ and enables robust measurements of weak absorption features that are difficult to constrain individually. Each spectrum is shifted into the rest frame using its systemic redshift and resampled onto a common rest-frame wavelength grid with the lowest resolution of the contributing spectra. We then continuum normalize the spectra by fitting the line-free regions and dividing the entire spectrum by the best-fit model, after which we combine them using inverse-variance weighting. This approach gives greater weight to well-constrained pixels and prevents a low-quality spectrum from disproportionately affecting the composite. Prior to considering the correlated noise, which we do below, the median SNR of our spectra is measured to be $\sim14$.

As a check that the composite is not driven by any single source, we repeat the combination using an unweighted median in place of the inverse-variance-weighted mean. The two composites agree to within $2\sigma$ in the rest-frame equivalent width of every transition used in this work and yield consistent doublet ratios, confirming that the measured line strengths are not dominated by any individual spectrum \extfigref{ext_fig:fig6}.

The number of contributing spectra decreases near the blue and red edges of the wavelength coverage. This does not affect the present analysis, however, because all absorption features used to measure the elemental abundances lie within the wavelength range covered by all four sources. We construct separate composite spectra for the G140M, G235M and G395M gratings rather than combining them, thereby preserving their different spectral resolutions and dispersions.

Resampling onto a common wavelength grid introduces covariance between neighboring pixels, causing the propagated pixel errors to underestimate uncertainties on integrated equivalent widths. We quantify this effect directly using nine line-free continuum windows between $1350$ and $3000$\,\AA. The measured scatter is $\sim30\%$ larger than expected from the propagated uncertainties, and neighboring pixels are also correlated because of the spectral resampling. Combining these effects we determine that the uncertainties on the EW measurements should be increased by a factor of $2.1$. We therefore conservatively inflate all equivalent-width uncertainties by a factor of two; these rescaled uncertainties are used throughout the column-density and curve-of-growth analyses.

This correction also determines which weak transitions are used quantitatively. We base the fiducial Fe\,{\sc ii} column density on the robust $\lambda1608$, $\lambda2344$, $\lambda2586$ and $\lambda2600$  transitions, while weaker Fe\,{\sc ii} features are retained only as consistency checks. Thus, marginal features introduced or amplified by correlated noise do not drive the inferred Fe abundance.

\begin{figure}
\centering
\includegraphics[width=1.0\linewidth]{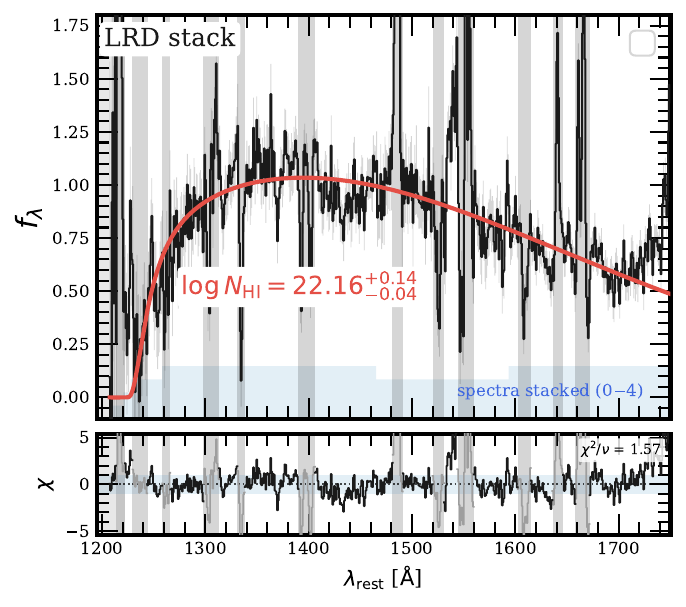}
\caption{Ly$\alpha$ damping wing measurement. The LRD stack is shown in black, with the best fit damped Ly$\alpha$ model in red. Gray bands mark emission and absorption lines that have been masked out of the fit. In the bottom panel we show a residual plot. The blue shaded band shows how many individual spectra contribute to the stack at each wavelength.}
\vspace{-6mm}
\label{ext_fig:fig_dla}
\end{figure}

\begin{figure}[t!]
\centering
\includegraphics[width=0.95\linewidth]{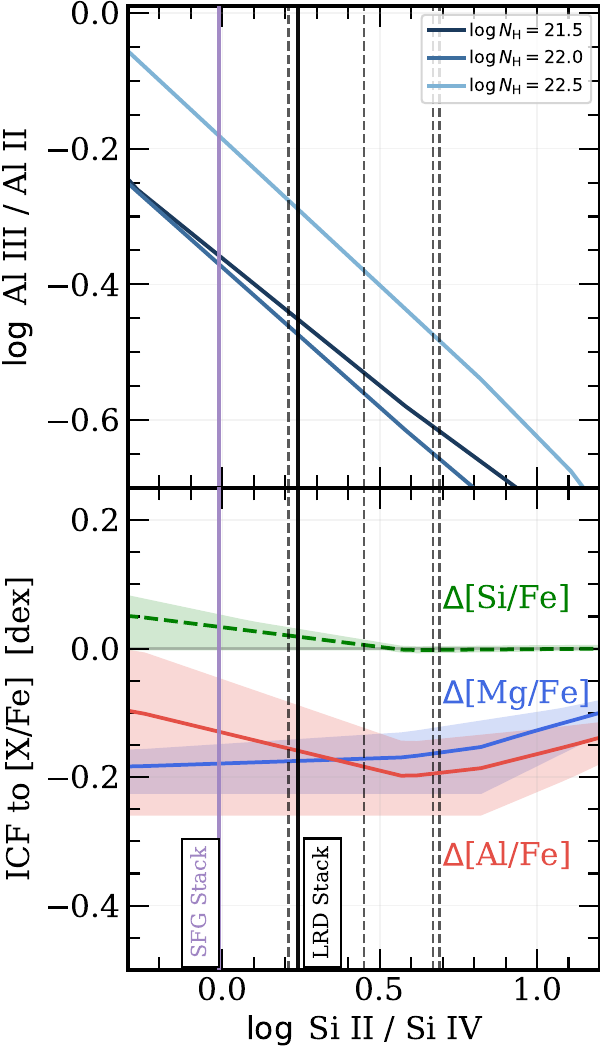}
\caption{\textbf{Ionization corrections anchored to the observed
silicon ion ratio.} \textbf{Top:} Predicted Al\,{\sc iii}/Al\,{\sc ii} ratio at different levels of ionization parametrized by the observed Si\,{\sc ii}/Si\,{\sc iv} ratio. LRD stack in black, individual LRDs in dashed gray, SFG stack in purple. \textbf{Bottom:} Ionization corrections to [Mg/Fe] (blue), [Al/Fe] (red) and [Si/Fe] (green) ratios given ionization.}
\label{ext_fig:fig4}
\end{figure}

\noindent
{\bf Star-Forming Galaxy Comparison Sample}
\\
\noindent
To establish whether the Mg--Al--Fe absorption pattern observed in the LRDs is specific to this population, we construct a comparison sample of ordinary star-forming galaxies at similar redshifts and ultraviolet luminosities.

Our comparison sample consists of 31 star-forming galaxies drawn from SPURS and selected to span $6.7\lesssim z\lesssim7.1$ and $M_{\rm UV}\sim-20\pm1$, closely matching the LRD sample. Their spectra are reduced, continuum-normalized, and combined using the same inverse-variance-weighted stacking procedure adopted for the LRDs, ensuring a direct and internally consistent comparison.

Mg\,{\sc ii} and Fe\,{\sc ii} absorption are common in the rest-frame ultraviolet spectra of star-forming galaxies, where they trace both outflowing and ambient interstellar gas along the line of sight to the stellar continuum. Their presence alone is therefore not indicative of unusual chemical enrichment. In \extfigref{ext_fig:fig2}, we directly compare the Al\,{\sc iii}, Mg\,{\sc ii}, and Fe\,{\sc ii} absorption in the LRD and matched star-forming-galaxy stacks. The contrast is striking. The Al\,{\sc iii}\,$\lambda1854$ equivalent width increases from $0.25\pm0.16$\,\AA\ in the SFG stack to $1.83\pm0.35$\,\AA\ in the LRDs, a factor of $\sim7$, while Al\,{\sc iii}\,$\lambda1862$ similarly increases from $0.14\pm0.15$\,\AA\ to $1.13\pm0.28$\,\AA. In contrast, Mg\,{\sc ii} absorption is stronger in the SFG stack: $W_r(2796)=3.73\pm0.16$\,\AA\ and $W_r(2803)=2.66\pm0.21$\,\AA, compared with $3.13\pm0.35$\,\AA\ and $1.87\pm0.28$\,\AA\ in the LRDs. Fe\,{\sc ii} shows only a moderate difference, with the clean $\lambda1608$ and $\lambda2344$ transitions stronger in the LRDs by factors of $\sim2$, while $\lambda2586$ is indistinguishable between the two populations. The aluminum excess is therefore directly apparent in the observed spectra relative to both Mg and Fe, independent of the detailed conversion from absorption strength to column density.

This comparison also provides an important empirical check on the abundance analysis. Even if moderate unresolved saturation affects the absolute ionic columns, it cannot readily explain why Al\,{\sc iii} changes so strongly between the two matched populations while Mg\,{\sc ii} remains comparable and Fe\,{\sc ii} changes only moderately.

To provide an independent low-$z$ comparison, we use published abundance measurements from the COS Legacy Archive Spectroscopic SurveY (CLASSY), a deep ultraviolet spectroscopic sample of nearby star-forming galaxies, to compare the location of local galaxies in [Al/Fe]--[Si/Fe] space with the LRD measurements \cite{meth:classy_main,meth:classy_main2,meth:classy1,meth:classy2,meth:classy3}.

\noindent
{\bf Column Density Measurements}
\\
\noindent
We estimate ionic column densities from the individual and stacked spectra using two complementary approaches. Our fiducial measurements use the apparent optical depth (AOD) method \cite{meth:aod}, in which the optical depth is integrated directly over the observed absorption profile. As an independent check, we also fit each resolved feature with a Gaussian optical-depth profile and derive the corresponding column density. The two approaches agree within their uncertainties for all reported transitions. For Fe\,{\sc ii}, where multiple transitions spanning a broad range in oscillator strength are available, we additionally test the inferred column density using a curve-of-growth analysis described in the section below. For each transition, we model the local continuum with a first-order polynomial fitted to line-free windows adjacent ($\pm2500$ km/s) to the feature of interest. The line-free continuum pixels are first binned by a factor of $10$--$20$, after which outlying binned points are iteratively rejected using a median absolute deviation (MAD) criterion. Because AOD column densities are sensitive to the adopted continuum level, the continuum windows and binning are adjusted individually for each feature to minimize systematic bias.

Systemic redshifts are defined by the rest-frame optical [O\,{\sc iii}]\,$\lambda\lambda4959,5007$ doublet. For each absorption feature, we define the line center at the position of minimum flux and record the corresponding velocity offset. The apparent optical depth is integrated over a velocity window of $\pm500$\,km\,s$^{-1}$ around this center, reduced where necessary for closely separated doublets to prevent overlap between the two components. The apparent optical depth is defined as
$\tau_a(v)=-\ln[I(v)/I_c(v)]$, where $I(v)$ and $I_c(v)$ are the observed and continuum flux densities at velocity $v$. The ionic column density is then
 \begin{equation*}
    N \;=\; \frac{m_e c}{\pi e^2}\,\frac{1}{f\lambda}
    \int \tau_a(v)\, \mathrm{d}v
    \;=\; \frac{3.768\times10^{14}}{f\,\lambda\,[\mathrm{\AA}]}
    \int \tau_a(v)\, \mathrm{d}v \;\; \mathrm{cm^{-2}},
\label{eq:aod}
\end{equation*}
where $m_e$ is the electron mass, $c$ is the speed of light, $e$ is the electron charge, $f$ is the oscillator strength of the transition, $\lambda$ is its rest wavelength, and $v$ is expressed in km\,s$^{-1}$. Uncertainties are propagated pixel by pixel from the error spectrum and the continuum fit. Atomic data for $f$ and $\lambda$ are taken from \cite{meth:morton03}. Rest-frame equivalent widths, $W_r$, are measured over the same velocity intervals. Features are treated as detections when $W_r$ SNR exceeds 2 and are otherwise reported as upper limits.

We flag obvious saturation directly from the observed profiles. A transition is treated as saturated when its trough reaches zero flux over multiple pixels or, for resolved doublets, when the relative strengths of the two components are inconsistent with the optically thin expectation (approximately 2:1 for the Mg\,{\sc ii} and Al\,{\sc iii} doublets used here). In these cases, the AOD column density is reported as a lower limit. More detailed tests for unresolved saturation using the Fe\,{\sc ii} curve of growth are presented below. The profile fits additionally provide the velocity centroids and intrinsic widths used in the subsequent kinematic analysis.

Several features require explicit decomposition before their column densities can be measured. Al\,{\sc ii}\,$\lambda1671$ is blended with O\,{\sc iii}]\,$\lambda\lambda1661,1666$ and He\,{\sc ii}\,$\lambda1640$ emission. We therefore fit the emission lines with Gaussian profiles, tying the O\,{\sc iii}] doublet ratio, while simultaneously modeling the Al\,{\sc ii} absorption. The column density is then measured from the emission-subtracted spectrum. 

Resonance doublets exhibiting P-Cygni profiles, specifically Si\,{\sc iv}\,$\lambda\lambda1394,1403$ and Mg\,{\sc ii}\,$\lambda\lambda2796,2803$, are decomposed into blueshifted absorption and near-systemic resonant-scattering emission (see \extfigref{ext_fig:fig_mgfit}). Although Si\,{\sc iv} can also exhibit P-Cygni profiles in the winds of young massive stars, the absorption observed here is kinematically coincident with the narrow absorption system traced by Al\,{\sc iii}, Mg\,{\sc ii}, Fe\,{\sc ii}, and H$\beta$ (see \extfigref{ext_fig:fig3}), favoring an origin in the dense gas within the LRD central source. Both members of each doublet share a common kinematic model, and the column density is measured from the emission subtracted absorption profile.

\noindent
{\bf Neutral Hydrogen Column Density}
\\
\noindent
We constrain the neutral hydrogen (H\textsc{i}) column density from the red damping wing of Ly$\alpha$ in the stacked spectrum. We restrict the measurement to the stack because no individual source combines sufficient continuum signal-to-noise with wavelength coverage close enough to the Ly$\alpha$ line core to constrain the damping profile independently.

The fiducial composite used for the metal-line analysis is continuum-normalized and therefore should not generally be used for this measurement, since this procedure can suppress the broad damping-wing shape. We instead reconstruct the stack from the original LRD spectra, normalizing each by its median flux density in the line-free windows $1340$--$1390$\,\AA\ and $1410$--$1465$\,\AA. The spectra are then resampled onto a common rest-frame wavelength grid and combined using inverse-variance weighting. No wavelength-dependent continuum correction is applied before stacking.

We fit the resulting spectrum with a curved power-law continuum attenuated by a damped Ly$\alpha$ absorption profile. The continuum is parameterized by a normalization $A$, slope $\beta$, and curvature $\gamma$, while the absorption is described by the full Lorentzian natural-broadening profile with $N_{\rm HI}$ as a free parameter. Allowing continuum curvature is important because otherwise intrinsic spectral curvature can be absorbed into the inferred damping opacity and bias $N_{\rm HI}$ high. We allow $|\gamma|\leq24$ and marginalize over both $\beta$ and $\gamma$ when deriving the $N_{\rm HI}$ posterior.

We fit the spectrum over $1222$--$1700$\,\AA, masking wavelength intervals containing neighboring emission/absorption features. The resulting fit gives
$\log\left[N_{\rm HI}/{\rm cm^{-2}}\right]=22.16^{+0.14}_{-0.04}$,
corresponding to a $3.5\sigma$ detection of the damping wing, with $\chi^{2}/\nu=1.57$ and best-fitting continuum parameters $\beta=-1.4$ and $\gamma=-13.0$ (\extfigref{ext_fig:fig_dla}).

This column is comparable to the highest neutral-hydrogen columns measured in $z>5$ galaxies with JWST, where damped Ly$\alpha$ systems reaching $\log N_{\rm HI}\simeq22.0$--$22.5$ have now been reported \cite{meth:heintz25,meth:hainline24,meth:deugenio24}. Similar columns are also inferred from narrow Balmer absorption in LRDs, with $\log N_{\rm H}\sim22$ \cite{meth:ji25}. By contrast, models that infer the gas column from the shape and strength of the Balmer break can require columns larger by $2$--$4$ dex \cite{meth:inayoshi25,meth:naidu25_bh*,meth:degraaff25_cliff}.

\noindent
{\bf Association of the Absorption with the Central Engine}
\\
\noindent
Tying the elemental abundances directly to the nature of LRDs hinges on establishing that the absorption arises in gas associated with the central engine. We test this by comparing the velocity offsets of the metal absorption lines with the blueshifted P-Cygni absorption in the broad H$\beta$ profile, which traces the dense gas surrounding the central engine \cite{meth:matthee26}. In the stacked spectrum, Al\,{\sc iii}, Mg\,{\sc ii}, Fe\,{\sc ii}, Si\,{\sc ii}, and Si\,{\sc iv} all exhibit absorption over a common velocity range of approximately $-140$ to $-200$\,km\,s$^{-1}$ relative to the systemic redshift defined by [O\,{\sc iii}] (\extfigref{ext_fig:fig3}). This absorption is coincident with the blueshifted P-Cygni component of the broad H$\beta$ profile. Because both the broad H$\beta$ emission and its associated absorption arise in the compact, dense gas surrounding the LRD central engine, the close kinematic correspondence strongly links the metal absorption to the same nuclear gas rather than to unrelated interstellar material in an extended host galaxy.

\noindent
{\bf Curve-of-Growth and Saturation Tests}
\\
\noindent
The AOD method can underestimate column densities when absorption is unresolved or strongly saturated. We test for this directly using Fe\,{\sc ii}, for which several transitions spanning a broad range in oscillator strength are detected. We fit the Fe\,{\sc ii} lines with Voigt profiles convolved with the empirical wavelength-dependent NIRSpec line-spread function \cite{meth:msaexp} and construct a curve of growth (COG) from the measured equivalent widths (\extfigref{ext_fig:fig_voigt}).

The Doppler widths measured directly from the resolved Fe\,{\sc ii} profiles are of order $b\sim200$\,km\,s$^{-1}$, as measured in the stack (\extfigref{ext_fig:fig_voigt}). These widths are independently supported by the other absorption tracers: H$\beta$, Mg\,{\sc ii}, Al\,{\sc iii}, Fe\,{\sc ii}, Si\,{\sc ii}, and Si\,{\sc iv} exhibit consistent velocity centroids and widths (\extfigref{ext_fig:fig2}, \extfigref{ext_fig:fig3}), indicating that they trace the same kinematic component. The absorption profiles are also broader than the instrumental line-spread function, demonstrating that the characteristic velocity width is resolved rather than imposed by the spectral resolution. We therefore use the line-profile measurements to constrain $b$ when evaluating the Fe\,{\sc ii} curve of growth.

At the measured $b$, the COG yields $\log N({\rm Fe\,II})\simeq15.1$, broadly consistent with the AOD column within the systematic uncertainties associated with continuum placement, spectral resolution and saturation. Thus, with the current low-resolution data quality, unresolved saturation does not have a quantifiable effect on the inferred column densities.

If $b$ is instead allowed to vary without an external kinematic constraint, the Fe\,{\sc ii} equivalent widths admit a second solution at $b\simeq68$\,km\,s$^{-1}$ and $\log N({\rm Fe\,II})\simeq16.7$ (\extfigref{ext_fig:fig_voigt}). This solution is driven primarily by the putative Fe\,{\sc ii}\,$\lambda1611$ feature with $SNR\sim2.2$, whose large equivalent width relative to its very small oscillator strength forces the curve onto the saturated part of the COG. However, such a narrow solution is inconsistent with the resolved widths measured directly from the Fe\,{\sc ii} profiles and with the broader, mutually consistent absorption seen in the other ions and in H$\beta$.

The high-column solution also produces a difficult abundance pattern. The same absorbing component has $\log N({\rm Mg\,II})\sim14$, so increasing Fe\,{\sc ii} to $\log N\simeq16.7$ would imply an exceptionally low Mg/Fe ratio, far more extreme than inferred from the fiducial AOD measurements. Conversely, if Mg were present at anything approaching a normal abundance relative to such a large Fe column, substantially stronger absorption would be expected in weak Mg\,{\sc ii} transitions, including Mg\,{\sc ii}\,$\lambda1240$, which is not detected. We therefore regard the low-$b$, high-$N$ branch as physically disfavored and adopt the COG solution constrained by the directly measured kinematic width.

We nevertheless cannot exclude unresolved narrow substructure within the broader absorption system at the present $R\sim1000$ spectral resolution. The measured $b$ therefore characterizes the bulk velocity distribution of the absorbing gas rather than necessarily a single physical cloud. Higher-resolution spectroscopy will ultimately be required to determine whether additional narrow saturated components are present. Within the available data, however, the agreement between the resolved line profiles, the COG and the AOD measurements suggest that no severe unresolved saturation drive the inferred column densities. Importantly, the qualitative aluminum excess does not depend on this conclusion, as it is also directly apparent in the matched LRD--SFG spectral comparison (\extfigref{ext_fig:fig2}).

Combining the adopted Fe\,{\sc ii} column with the neutral-hydrogen column measured from the Ly$\alpha$ damping wing gives a gas-phase metallicity of ${\rm [Fe/H]}=-2.56\pm0.30$. The narrow absorption component surrounding the LRD central source is therefore extremely metal poor, despite exhibiting the strong aluminum enhancement described above.

\noindent
{\bf CLOUDY Modeling and Ionization Corrections}
\\
\noindent
Converting the measured ionic columns into elemental abundance ratios requires particular care for aluminum. Al\,{\sc ii}\,$\lambda1670$ is a singlet, and therefore provides no independent transition with which to diagnose saturation. Its column density cannot be measured reliably, so our aluminum constraint relies primarily on the Al\,{\sc iii}\,$\lambda\lambda1854,1862$ doublet. However, Al\,{\sc iii} traces a higher ionization state than Fe\,{\sc ii} and Mg\,{\sc ii}: the ionization potentials required to produce these ions are 18.8, 7.9, and 7.6\,eV, respectively \cite{meth:berg21}. We must therefore constrain the ionization state of the absorbing gas before comparing Al\,{\sc iii} with the low-ionization species.

We use the Si\,{\sc ii}/Si\,{\sc iv} column-density ratio as an empirical tracer of the ionization state. Silicon is the only element in our line list with well-detected transitions spanning two widely separated ionization stages. Both stages also provide internal checks for saturation: Si\,{\sc ii} through the $\lambda1260$, $\lambda1304$, and $\lambda1527$ transitions, which span a broad range of oscillator strengths, and Si\,{\sc iv} through the resolved $\lambda\lambda1394,1403$ doublet. We translate the observed silicon ratio into an aluminum ionization correction using a grid of photoionization models computed with \textsc{Cloudy} \cite{meth:ferland17}. The models consist of plane-parallel slabs illuminated by a $45{,}000$\,K blackbody, chosen to represent the stellar-like ultraviolet continuum of the LRDs. We adopt a range of metallicities from $0.003-0.1\,Z_\odot$ with solar-scaled abundance ratios. The grid spans ionization parameters of $-3.0\leq\log U\leq-0.5$ in steps of 0.25 and stopping hydrogen columns of $21.0\leq\log N_{\rm H}\leq22.5$, consistent with the measured $N_{\rm HI}$. Each model is iterated to convergence and continued to a stopping temperature of 100\,K, ensuring that the full low-ionization zone is included. We then extract the integrated column densities of the relevant ionization stages of Mg, Al, Si, and Fe. For each source, the measured Si\,{\sc ii}/Si\,{\sc iv} ratio selects the subset of models consistent with its ionization state. From these models, and assuming $N_{\rm HI}\sim22$, we derive the correction required to place the observed Al\,{\sc iii} column on the same abundance scale as Mg\,{\sc ii} and Fe\,{\sc ii} before forming the elemental ratios.

We also use the same model grid to test whether ionization could mimic the observed magnesium depletion or aluminum enhancement by shifting a substantial fraction of either element into unobserved ionization stages. We track the ionic fractions, $X^{+}/X_{\rm total}$, as a function of ionization parameter, or equivalently Si\,{\sc ii}/Si\,{\sc iv}. Across the full range of models consistent with the observed silicon ratios, we find that [Al/Fe] can be reduced by at most $\sim$0.15\,dex, while [Mg/Fe] cannot be increased at any point in the grid: Mg\,{\sc ii} and Fe\,{\sc ii} have closely matched ionization potentials, so their ionic fractions track each other and the correction never exceeds $-0.2$ dex. The magnesium deficit is therefore robust to ionization, and the aluminum enhancement can only be modestly softened, not removed. The abundances we report in \figref{fig:fig2}, have all been corrected by using the values shown in \extfigref{ext_fig:fig4}.

\begin{figure}[t!]
\centering
\includegraphics[width=0.9\linewidth]{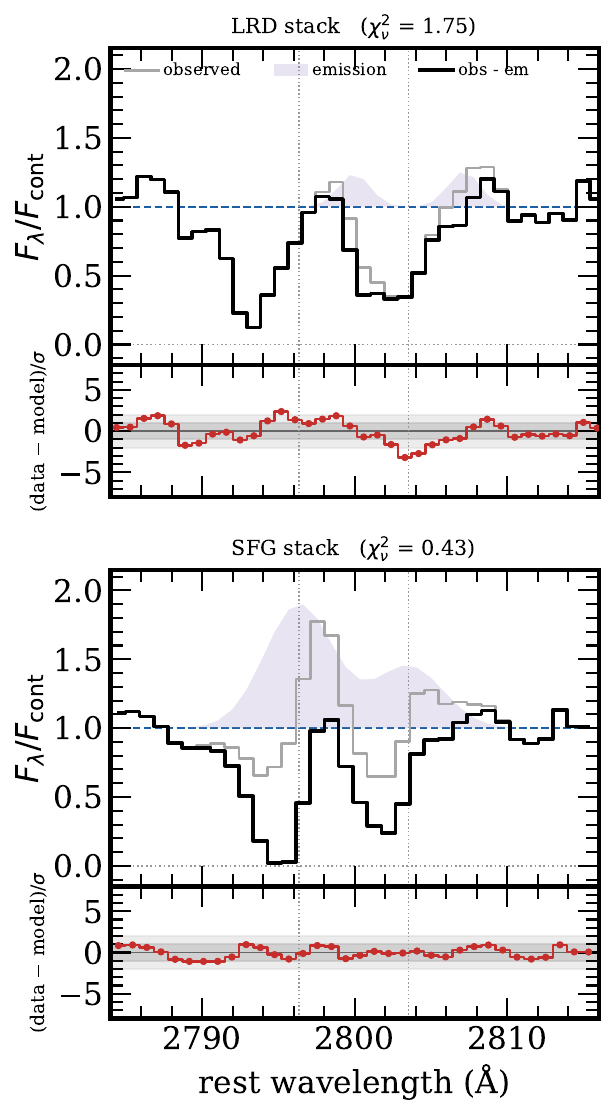}
\caption{Decomposition of the Mg\,{\sc ii}~$\lambda\lambda2796,2803$ P~Cygni profile
in the LRD composite (\textbf{top}) and the SFG composite (\textbf{bottom}). Grey: observed continuum-normalized stack; shaded: best-fit emission component; black: the emission-subtracted profile used to measure the absorption. Dotted
vertical lines mark the doublet rest wavelengths. Lower panels show residuals of the full model.}
\vspace{-6mm}
\label{ext_fig:fig_mgfit}
\end{figure}

\begin{figure}[t!]
\centering
\includegraphics[width=1.0\linewidth]{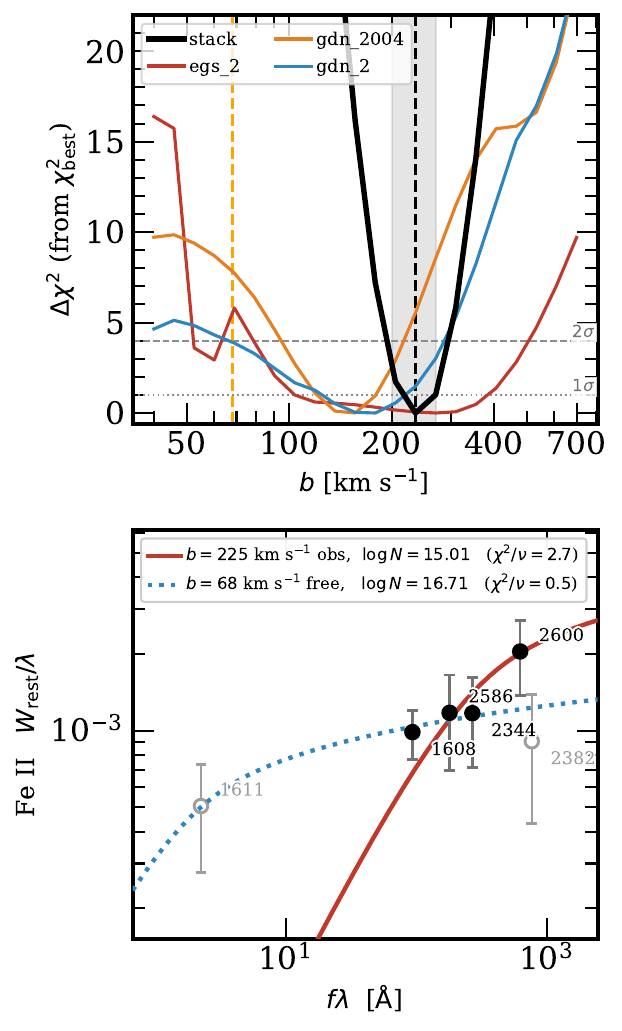}
\caption{\textbf{Curve-of-growth constraints on the Fe\,{\sc ii} column density.}
\textbf{Top:} Change in $\chi^2$ as a function of Doppler parameter $b$ for the Fe\,{\sc ii} curve-of-growth fits to the stacked spectrum (black) and individual LRDs (colored curves). The Fe\,{\sc ii} transitions are resolved and are generally fit with $b\sim180-250$ km/s. The solution with a free $b$ value (orange dashed) is $\sim 5\sigma$ away from the measured $b$ for the stack (black dashed) of $225\pm35$ km/s. \textbf{Bottom:} Rest-frame equivalent widths of the Fe\,{\sc ii} transitions as a function of $f\lambda$. The dotted blue curve shows the free-$b$ solution, while the solid red curve fixes the Doppler parameter to the values measured directly from the Fe\,{\sc ii} lines. Lines with SNR$<3$ are grayed out.}
\vspace{-6mm}
\label{ext_fig:fig_voigt}
\end{figure}

\begin{figure}[t!]
\centering
\includegraphics[width=\columnwidth]{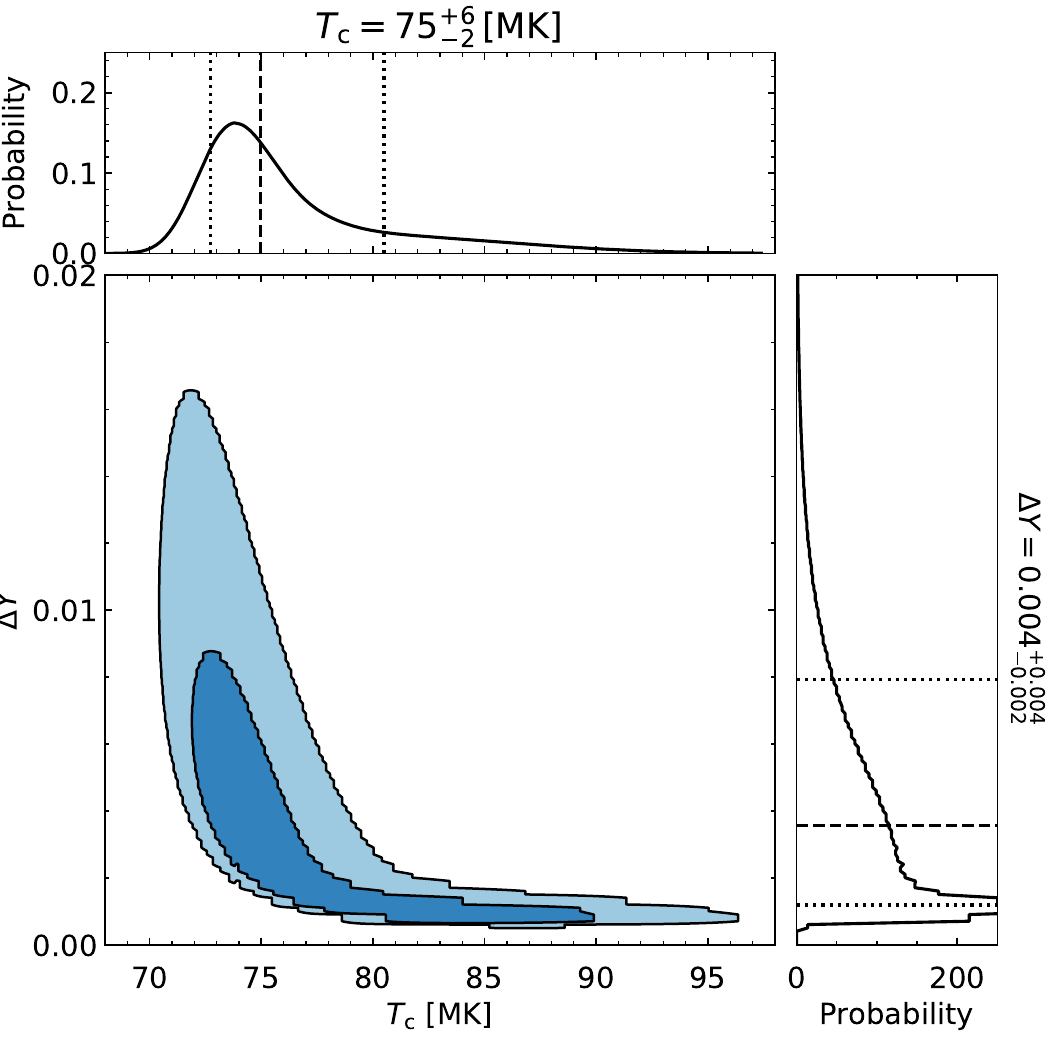}
\caption{\textbf{Best Fit Nucleosynthesis Parameters} 
Likelihood contours ($1\sigma$ dark and $2\sigma$ light) of the nucleosynthesis calculations with  {\sc pynucastro} that reproduce the average abundances of [Mg/Fe], [Al/Fe] and [Si/Fe]. The marginalized posteriors for $T_{\rm c}$ and $\Delta Y$ are shown in the top and right panels, respectively.  }
\vspace{-6mm}
\label{ext_fig:fig5}
\end{figure}

\noindent
{\bf Nucleosynthesis Comparison}
\label{sec:tracks}
\\
\noindent
The broader family of light-element abundance anomalies produced by proton-capture processing is well established in spectroscopic studies of Galactic globular clusters \cite{meth:gratton04,meth:carretta09}.
We run nucleosynthesis calculations of hydrogen burning  at different temperatures with {\sc pynucastro} \cite{meth:smith2023,meth:pynucastro_v2.8.0}. We use a net with 43 species from H up to Ca, and all relevant reactions for hydrogen burning, including the CNO-cycle, the NeNa-cycle and the MgAl-cycle. For the initial abundances we scale solar abundances from \cite{meth:lodders21} to [Fe/H]$=-2.5$, matching the metallicity measured for the LRD stack, and an initial helium mass fraction of $Y=0.248$. We fix the density to 0.1 g/cm$^3$, appropriate for the centers of SMSs\cite{meth:prantzos17}. All calculations are run until the helium mass fraction reaches $Y=0.99$. 
 
For illustrative purposes (Figs.~\ref{fig:fig2} and \ref{fig:fig4}) we run a grid with temperature between 60 MK and 90 MK with steps of 5 MK for $\afe = 0$. In addition to this coarse grid, we also compute a fine grid to determine at high resolution the likelihood of the average abundances of the four LRDs  (${\rm [Mg/Fe]} = -1.08  \pm 0.14, {\rm [Al/Fe]} = 0.90 \pm 0.21, {\rm [Si/Fe]} = -0.05\pm 0.09$) for different $T_{\rm c}$ and $\Delta Y$. The obtained parameters are $T_{\rm c} = 75^{+6}_{-2}\,{\rm MK}$, $\Delta Y = 0.004^{+0.004}_{-0.002}$ (see \extfigref{ext_fig:fig5}), which leads to the recovered observables: ${\rm [Mg/Fe]} = -1.08^{+0.16}_{-0.12}, {\rm [Al/Fe]} = 1.01^{+0.12}_{-0.09}, {\rm [Si/Fe]} = 0.04^{+0.04}_{-0.02}$. The shape of the contours show that  $T_{\rm c}$ and $\Delta Y$ are somewhat degenerate. This is because the Mg-Al chain is relative slow (compared to the CNO-cycle and NeNa-cycle). Because $\Delta Y$ is a proxy for time, one can obtain more Mg-depletion and Al-enhancement with either a higher temperature, or with a lower temperature but at a later time. 

The recovered parameters can be used to infer a mass of a polluter. The dependence of $T_{\rm c}$ on stellar mass ($m$) and $\Delta Y$ found in stellar models presented in \cite{meth:prantzos17} can be  approximated by a simple linear bi-variate relation $T_{\rm c}(m,\Delta Y)$ \cite{meth:gieles25} (their equation 25). Using this, we find for the mass of the polluter: $\log_{10}m = 4.0^{+0.7}_{-0.4}$. At sufficiently high masses, the subsequent evolution of supermassive stars is affected by the general-relativistic instability \cite{meth:fowler66,meth:nandal26_puls}.

To assess the possible impact of unresolved saturation in Mg\,{\sc ii}, we repeat the inference after increasing the measured ${\rm [Mg/Fe]}$ abundance by 0.5 dex. The resulting posterior distributions in $T_{\rm c}$ and $\Delta Y$ become substantially broader, but remain consistent with the fiducial solution within the correspondingly larger uncertainties. This demonstrates that the inference is not driven solely by the extreme Mg depletion: the strong Al enhancement relative to ordinary star-forming galaxies which we show in \extfigref{ext_fig:fig2} provides an independent constraint and remains the principal driver of the hot hydrogen-burning interpretation.

\begin{figure*}
\centering
\includegraphics[width=1.0\linewidth]{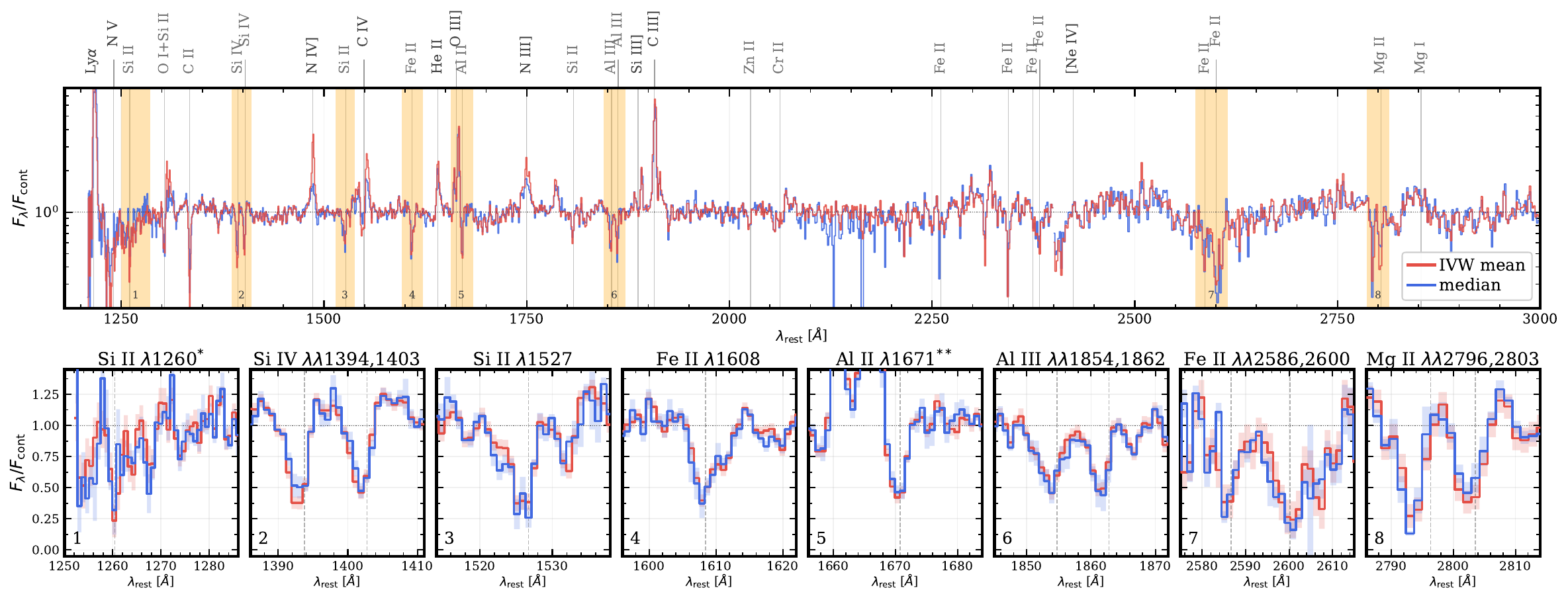}
\caption{\textbf{Inverse-variance vs median diagnostic.} Top panel shows the comparison between the inverse-variance (IVW; red) and median (blue) G140M+G235M spectra in our region of interest. Bottom panel shows the zoomed-in comparison of the absorption lines between different stacks.}
\vspace{-6mm}
\label{ext_fig:fig6}
\end{figure*}

\begin{figure*}
\centering
\includegraphics[width=.8\linewidth]{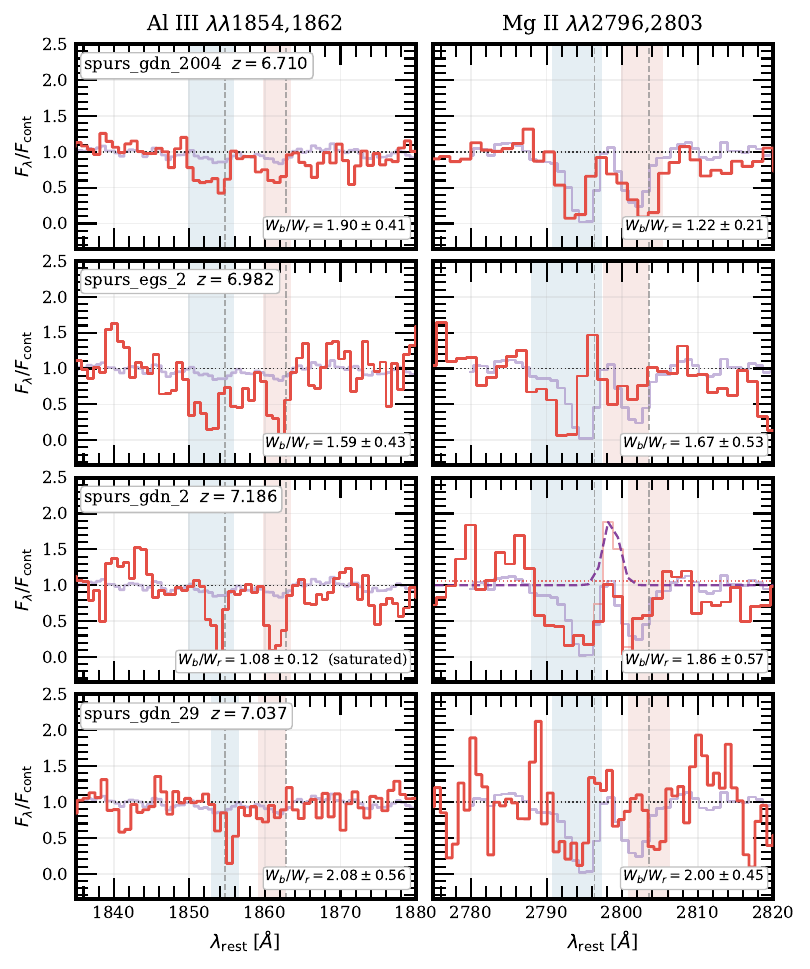}
\caption{\textbf{Al\,{\sc iii} and Mg\,{\sc ii} doublet ratios for individual LRDs.} Continuum-normalized profiles of the Al,{\sc iii},$\lambda\lambda1854,1862$ (left) and Mg,{\sc ii},$\lambda\lambda2796,2803$ (right) doublets for each of the four LRDs. In the background, the SFG stack is shown for comparison. The ratio of the rest-frame equivalent widths of the blue and red doublet members, $W_{\rm b}/W_{\rm r}$, is indicated in each panel; the optically thin expectation is approximately two, while ratios approaching unity indicate increasing saturation. Transitions identified as saturated are marked explicitly and their apparent-optical-depth column densities are treated as lower limits. Vertical lines show the line location at $V=0$. While our primary abundance constraints rely on the stacked spectrum, the generally resolved doublet structure provides a source-by-source check on saturation.}
\label{ext_fig:fig8}
\end{figure*}

\begin{sidewaystable*}[h]
\centering
\caption{\textbf{Adopted ionic column densities and abundance ratios.}
Column densities are $\log_{10}[N(\mathrm{ion})/\mathrm{cm}^{-2}]$.
Unless otherwise stated, column densities were measured with the apparent-optical-depth method from the transition with the highest continuum SN.  Abundance ratios are relative to solar \cite{asplund09} and include
the Al\,\textsc{iii}$\rightarrow$Al\,\textsc{ii} correction and the ICF described in
Methods, directly corresponding to the values plotted in \figref{fig:fig2}.
The gas-phase metallicity [Fe/H] is quoted only for the LRD stack, for which the
neutral hydrogen column is constrained by the damped Ly$\alpha$ profile.}
\label{tab:ionic_columns}
\scriptsize
\setlength{\tabcolsep}{3.5pt}
\begin{tabular}{lccccccccccc}
\toprule
SPURS ID & $z_{\rm spec}[\rm{O\,\textsc{iii}}]$ &
$\log N(\mathrm{Mg\,\textsc{ii}})$ &
$\log N(\mathrm{Al\,\textsc{ii}})$ &
$\log N(\mathrm{Al\,\textsc{iii}})$ &
$\log N(\mathrm{Fe\,\textsc{ii}})$ &
$\log N(\mathrm{Si\,\textsc{ii}})$ &
$\log N(\mathrm{Si\,\textsc{iv}})$ &
[Mg/Fe] & [Al/Fe] & [Si/Fe] & [Fe/H] \\
\midrule
SFG stack & $\cdots$ & $14.40\pm0.35$ & $13.05\pm0.03$ & $13.56\pm0.06$ & $14.75\pm0.07$ & $14.66\pm0.10$ & $14.67\pm0.06$ & $-0.62\pm0.36$ & $+0.08\pm0.09$ & $-0.11\pm0.12$ & $\cdots$ \\
LRD stack & $\cdots$ & $14.05\pm0.18$ & $13.81\pm0.03$ & $14.36\pm0.06$ & $15.01\pm0.15$ & $15.03\pm0.10$ & $14.79\pm0.06$ & $-1.22\pm0.23$ & $+0.69\pm0.16$ & $-0.01\pm0.18$ & $-2.56\pm0.30$ \\
\midrule
gdn\_2 & 7.1862 & $14.33\pm0.28$ & $13.93\pm0.12$ & $>14.69$ & $14.93\pm0.21$ & $14.97\pm0.14$ & $14.76\pm0.17$ & $-0.87\pm0.35$ & $>+1.10$ & $+0.01\pm0.25$ & $\cdots$ \\
gdn\_29 & 7.0372 & $14.02\pm0.33$ & $13.96\pm0.19$ & $14.22\pm0.23$ & $14.93\pm0.21$ & $15.00\pm0.14$ & $14.31\pm0.17$ & $-1.16\pm0.39$ & $+0.78\pm0.31$ & $+0.02\pm0.25$ & $\cdots$ \\
gdn\_2004 & 6.7102 & $13.98\pm0.13$ & $13.94\pm0.05$ & $14.33\pm0.08$ & $15.02\pm0.10$ & $15.25\pm0.11$ & $14.80\pm0.16$ & $-1.30\pm0.16$ & $+0.72\pm0.13$ & $+0.19\pm0.15$ & $\cdots$ \\
egs\_2 & 6.9816 & $14.01\pm0.40$ & $14.26\pm0.27$ & $14.69\pm0.14$ & $15.20\pm0.20$ & $15.07\pm0.20$ & $14.40\pm0.20$ & $-1.44\pm0.45$ & $+0.97\pm0.24$ & $-0.18\pm0.28$ & $\cdots$ \\
\bottomrule
\end{tabular}
\end{sidewaystable*}

\clearpage

\end{methods}

{\renewcommand\refname{References for Methods}%
\providecommand{\noopsort}[1]{}

}

\begin{addendum}

\item[Acknowledgments]
VK acknowledges support from the University of Texas at Austin Cosmic Frontier Center.  Support for this work was provided by NASA
through the NASA Hubble Fellowship grant HST-HF2-51607.001-A awarded by the Space Telescope Science Institute, which is operated by the Association of Universities for Research in Astronomy, Incorporated, under NASA contract NAS5-26555. These observations are associated with program GO 9214 (PIs: C. Mason \& D. Stark). This work is based on observations made with the NASA/ESA/CSA \textit{James Webb Space Telescope}, obtained at the Space Telescope Science Institute, which is operated by the Association of Universities for Research in Astronomy, Incorporated, under NASA contract NAS5-03127. The JWST data presented in this article were obtained from the Mikulski Archive for Space Telescopes (MAST) at the Space Telescope Science Institute.

\item[Author Contributions] All authors contributed to aspects of the analysis and to the writing of the manuscript.

\item[Author Information] Correspondence and requests for materials should be addressed to VK (vkokorev@utexas.edu).

\item[Code Availability] All results presented may be reproduced with the open access reduced data described above and using the following publicly available software: \texttt{msaexp}, \texttt{grizli}, \texttt{pynucastro}, \texttt{astropy}, \texttt{Cloudy}, \texttt{SpectRes},  

\end{addendum}

\begin{affiliations}

\item Department of Astronomy, The University of Texas at Austin, Austin, TX 78712, USA

\item Cosmic Frontier Center, The University of Texas at Austin, Austin, TX 78712, USA

\item Institute for Astronomy, University of Hawai‘i, 2680 Woodlawn Drive, Honolulu, HI 96822, USA

\item ICREA, Pg. Llu\'is Companys 23, E-08010 Barcelona, Spain

\item Institut de Ci\'encies del Cosmos (ICCUB), Universitat de Barcelona (UB), c. Mart\'i i Franqu\'es, 1, E-08028 Barcelona, Spain

\item Institut d’Estudis Espacials de Catalunya (IEEC), Edifici RDIT, Campus UPC, E-08860 Castelldefels (Barcelona), Spain

\item Department of Astrophysical Sciences, 4 Ivy Lane, Princeton University, Princeton, NJ 08540

\item David A. Dunlap Department of Astronomy and Astrophysics, University of Toronto, 50 St. George Street, Toronto, Ontario, M5S 3H4, Canada

\item Max Planck Institute for Astronomy, Heidelberg, Germany

\item Center for Astrophysics, Harvard and Smithsonian, 60 Garden Street, Cambridge, MA 02138, USA

\item Institute of Science and Technology Austria (ISTA), Am Campus 1, 3400 Klosterneuburg, Austria

\item Center for Interdisciplinary Exploration and Research in Astrophysics (CIERA), Northwestern University, 1800 Sherman Avenue,
Evanston, IL 60201, USA

\item Department of Astronomy, New Mexico State University, 1320 Frenger Mall, Las Cruces, NM 88003-8001, USA

\end{affiliations}

\end{document}